\documentclass[11pt,a4paper]{article}
\usepackage{jheppub_kim}
\usepackage{rotating}
\usepackage{graphicx,epsfig}
\usepackage{amsmath}
\usepackage {amssymb}
\usepackage{subfigure}
\usepackage{adjustbox}
\usepackage{multirow}

\usepackage{relsize}

\usepackage{array,multirow}
\usepackage{soul}
\usepackage{subfigure}
\usepackage{dsfont}
\usepackage{hyperref}
\usepackage{txfonts}
\usepackage{newlfont}
\usepackage{times}

\renewcommand{\figurename}{{\bf Fig.}}
\renewcommand{\tablename}{{\bf Tab.}}

\title{\boldmath  Strangeness Enhancement at LHC Energies using the thermal model and EPOSLHC event-generator }

\author[a]{Mahmoud Hanafy,}
\author[b]{Omnia S.A. Qandil,}
\author[a]{Asmaa G. Shalaby}

\affiliation[a]{Physics Department, Faculty of Science, Benha University, Benha 13518, Egypt.}
\affiliation[b]{Mathematics and Theoretical Physics Department, Atomic Energy Authority, Cairo, Egypt.}

\emailAdd{Mahmoud.Nassar@fsc.bu.edu.eg}
\emailAdd{omnia.shaker@gmail.com}

\abstract{ The strangeness enhancement signature of QGP formation at LHC energies is carefully tackled in the present study. Based on HRG, the particle ratios of mainly strange and multi-strange particles are studied at energies from lower $\sqrt{s}\sim $ 0.001 up to 13 TeV. The strangeness enhancement clearly appeared at more higher energies, and the ratios are confronted to the available experimental data. The particle ratios are also studied using the Cosmic Ray Monte Carlo (CRMC) interface model with its two different event generators namely; EPOS $1.99$ and EPOSlhc which show a good agreement with the model calculations at the whole range of the energy. We utilize  to produce some ratios. EPOS $1.99$ is used to estimate particle ratios at lower energies from AGS up to the Relativistic Heavy Ion Collider (RHIC) while EPOSlhc is used at LHC energies. The production of kaons and lambda particles is studied in terms of the mean multiplicity in p-p collisions at energies ranging from 4 - 26 GeV. We find that both HRG model and the used event generators, EPOS $1.99$ and EPOSlhc, can describe the particle ratios very well. Additionally, the freeze-out parameters are estimated for different collision systems, such as p-p and Pb-Pb, at LHC energies using both models.}


\begin{document}
\maketitle
\flushbottom

\section{Introduction}
One of the most important signatures of the phase transition between the hadronic matter "confined phase" to the quark-gluon plasma (QGP) "deconfined phase" is the strangeness enhancement, in other words, the production of strange particles \cite{Tiwari2014}. The abundance of "s" quark is a useful tool to analyze the heavy-ion and proton-proton collisions. Additionally, abundance of strangeness is considered as an abundance in the degree of freedom.

The strangeness enhancement is combined with gluons existence in QGP, in which the gluon dissolves to a pair of strange quarks rapidly \cite{Koch2017}. 
 An early study explored the phase transition from hadronic matter to QGP \cite{Rafelski1980}, and postulated the idea of chemical and thermal equilibrium which, in turn, developed the explanation of thermodynamics at high temperature and various kinds of chemical potentials.

It is noted that QGP comprises an equal number of strange and anti-strange quarks \cite{Rafelski2015}. Therefore, the density of strange quarks raises, and more multi-strange hadrons are produced. This occurs during the hadronization process \cite{Rafelski1982}. 
Recently, a good review \cite{Rafelski2020} handled the strangeness production as a signature for QGP formation. The theoretical and experimental procedures are discussed besides tackling the non-strange signature of QGP such as $J/\Psi$ suppression.

In the present work, the Hadron Resonance Gas (HRG) model is utilized as a powerful tool to analyze the production of hadrons resulting from various heavy-ion experiments such as AGS, SPS \cite{AGS_SPS1, AGS_SPS2}, RHIC \cite{RHIC, RHIC_more} and LHC \cite{Braun99}. This is in addition to the previous work that investigated RHIC, LHC and NICA energies \cite{Our2015,Our2016}.

An earlier work \cite{Tiwari2014} studied the strange and non-strange production in the framework of excluded volume model which  fits good with the experimental data. The strange to non-strange ratios are analyzed, in particular the kaon to pion and $\Lambda$ to pion ratios in a canonical ensemble \cite{Oeschler16}. The results showed an effect of the system size, and as a consequence the peak (horn) of such ratios are noticed at different energies. Another interesting work of the system-size dependence of hadrochemistry is applied \cite{Vovechenko19}. The enhancement of multi-strange hadrons in
high-multiplicity pp collisions \cite{Nature_Alice}.

In addition, the particles $\phi(1020), K_{s}^{0*}(892)$ have a vital role in heavy-ion collisions \cite{PDG2014} throughout the evolution process. This could be attributed to their short life times ($4.16 \pm0.05 \hspace{0.1cm}fm/c$, $46.3 \pm 0.4$ $fm/c$) respectively, that facilitate analyzing the system at various times. The enhanced contribution of these particles is essential due to the strangeness enhancement.

The particles $\phi(1020), K_{s}^{0*}(892)$ production are studied  \cite{ALICE2017} in lead-lead (Pb-Pb) and proton-proton (p-p) collisions at nucleon center of mass energy ($\sqrt{s_{NN}}$=$2.76$ TeV). For a good knowledge a pedagogic review in strangeness enhancement and papers therein \cite{Rafleski08}.

The main target of the present work is to investigate the strangeness enhancement in terms of various particle ratios such as $k^{+} / \pi^{+}$, $k^{-} / \pi^{-}$, $\pi^{-} / \pi^{+}$, $k^{-} / k^{+}$, $\bar{p}/\pi^{-}$, $\Lambda / \pi^{-}$, $\overline{\Lambda} / \Lambda$, $\overline{\Omega} / \Omega$, and $\overline{\Xi^{+}} / \Xi^{-}$, and strange and multi-strange particles such as $\phi(1020) / k_{s}^{0*}(892)$, $\overline{\Sigma}^{*+8} / \Sigma^{*+8}$, $\overline{\Sigma}^{*08} / \Sigma^{*08}$, $\overline{\Sigma}^{*-8} / \Sigma^{*-8}$, $\overline{\Xi}^{*08} / \Xi^{*08}$, and $ \overline{\Xi}^{*-8} / \Xi^{*-8}$ from low to high energies using different models namely; HRG, EPOS $1.99$ and EPOSlhc. Both event generators, EPOS $1.99$ and EPOSlhc, are executed through the CRMC interface model to produce the above mentioned  particles for an ensemble of $100,000$ events where the fusion option is turned on. The production of kaons and lambda particles is studied in terms of the mean multiplicity in p-p collisions at energies ranging from 4 - 26 GeV. First, we have used both event generators to produce well-identified particles and comparing the obtained results with the available experimental data. This encourage us to use it for a production of particles which haven't experimental data. Also, the freezeout parameters, i.e., the temperature ($T_{ch}$) and baryon chemical potential ($\mu_{B}$), are estimated as a result of fitting the obtained results from the HRG model of a combination of the used particle ratios with both LHC and EPOSlhc results. The obtained values of $T_{ch}$ and $\mu_{B}$ is compared to that presented in Ref. \cite{Rathexp_fit}.

The present study is organized as follows:
in Section \ref{model}, the main equations of the hadron gas model are discussed. A general introduction about the event generator is presented in section \ref{simulation}. Section \ref{results_dis} presents the obtained results. Finally, the conclusion is represented in Section \ref{conclusion}. 

\section{Formalism}\label{model}
 In the present work, the grand canonical ensemble (GCE) is used in the framework of the HRG model. In GCE ensembles, the energy exchanges freely with the surrounding medium. So that, the number of particles is no longer fixed. Such a system posses thermodynamic properties which can obtained from the GCE partition function. The GCE has rigorous conserved quantum numbers such as the charge, strangeness and baryon quantum numbers. Thus, the GCE partition function is defined as follows \cite{Redlich02,Redlich03}
 
\begin{equation}
Z(T,V,\mu_{Q})= Tr[exp(-\beta (H-\sum_{i} \mu_{Q_{i}} Q_{i}))],
\label{eq:1}
\end{equation}
Where H is the Hamiltonian, $Q_{i}$ is the different conserved charges, $\mu_{Q_{i}}$ are the corresponding chemical potentials, and $\beta=\frac{1}{T}$ in natural units ($\hbar = c = k_{B} = 1$). The Hamiltonian in HRG includes all the degree of freedom. Then the partition function in the hadron resonance gas can be written as a sum of partition functions of hadrons and resonances as follows \cite{Redlich02,Redlich03}
\begin{equation}
 \ln Z (T,V,\vec{\mu}) =  \sum_{i} \ln Z_{i}(T,V,\vec{\mu})=\pm \sum_{i} \frac{V g_{i}}{(2\pi)^{2}} \int_{0}^{\infty}  k^{2} dk \ln[1\pm \lambda_{i}\exp(-\beta \hspace{0.1cm} \varepsilon_{i})],
\label{eq:2}
 \end{equation}
where the $\pm$ signs refer to fermions and bosons, respectively, $\varepsilon_{i}= \sqrt{k^{2}+m_{i}^{2}}$ with $m_{i}$ is the mass of $" \textit{i} "$ particle and, $\lambda_{i}(T,\vec{\mu})$ is the fugacity factor and is given by \cite{Redlich02,Redlich03}.    

\begin{equation}
\lambda_{i}(T,\vec{\mu})= exp\left( \frac{\mu_{s} S_{i}+ \mu_{q} Q_{i}+ \mu_{B} B_{i}}{T}\right),
\label{eq:3}
\end{equation}

 with $ \mu_{s}, \mu_{q}$ and $ \mu_{B} $ are the strange, quark and baryon chemical potentials respectively, and $S_{i}, Q_{i}, B_{i}$ are the corresponding quantum numbers for particle species "$i$". These quantities should fulfill the conservation laws such as strangeness, $\sum_{i} S_{i} \hspace{0.03cm}N_{i}=0$, and charge and baryon number, $\frac{\sum_{i} Q_{i} \hspace{0.03cm}N_{i}}{\sum_{i} B_{i} \hspace{0.03cm}N_{i}}= \frac{Z}{A}$, where Z, A are the atomic number and mass number of the colliding nuclei, respectively.
 The integration in eq. (\ref{eq:2}) has been performed over "$k$" resulting in Bessel function $K_{2}$ \cite{Redlich03} 
\begin{eqnarray}
 \ln Z_{i}(T,V,\vec{\mu}) = \frac{V T g_{i}}{(2\pi)^{2}} \sum_{n=1}^{\infty} \frac{(\pm 1)^{n+1}}{n^{2}} \lambda_{i}^{n} m_{i}^{2} K_{2}(n m_{i} \beta),
 \label{eq:4}
\end{eqnarray}

Therefore, the thermodynamic quantities can be obtained from Eq.(\ref{eq:4}). Then the number density of particles is given by \cite{Redlich03}

\begin{equation}
n_{i}(T,\vec{\mu})= \frac{<N_{i}>}{V}=\frac{ T g_{i}}{(2\pi)^{2}} \sum_{n=1}^{\infty} \frac{(\pm 1)^{n+1}}{n^{2}} \lambda_{i}^{n} m_{i}^{2} K_{2}(n m_{i} \beta). 
\label{eq:5}
\end{equation}

where $<N_{i}>$ is the average number of the particles. In order to include all hadrons with their resonance decay, the average number can be rewritten as

\begin{equation}
<N_{i}>_{total} = <N_{i}>^{th} + \sum_{j}\textbf{Br} (j\rightarrow i)<N_{j}>^{th,R}.
\label{eq:6}
\end{equation}
where the first term represents the average number of thermal particles of species $i$ and the second term represents all resonance contributions to the particle multiplicity of species $i$ with "\textbf{Br}" stands for the branching ratio for the decay from particle ($j\rightarrow i$). All particles ratios are calculated using Eq. (\ref{eq:5}).

\section{Cosmic Ray Monte Carlo (CRMC)} \label{simulation}

CRMC is an interface which gives accessing to various Monte Calro event generators such as EPOS 1.99, EPOSlhc, SIBYLL 2.1/2.3, QGSJet 01/II.03/II.04 \cite{Werner2006,Pierog2013,Urlich_site}. CRMC provides a full background description taking into account the produced diffraction. It is built on various types of interactions which are depending on the Gribov-Regee model such as EPOS $1.99$ and EPOSlhc.

EPOS $1.99$ and EPOSlhc are designed to explain both cosmic and non-cosmic air showers and could be used to describe data produced from various collision systems such as proton-proton "p-p" or proton-nucleus "p-A" or deutron-nucleus "d-Au" gold. Others in \cite{Werner2006} presented a phenomenological approach based on the parton model trying to understand different experiments by a unified approach. They introduced EPOS, which stands for \textbf{E}nergy-conserving quantum mechanical multiple scattering approach, based on \textbf{P}artons (parton ladders), \textbf{O}ff-shell remnants, and \textbf{S}plitting of parton ladders \cite{Werner2006}. EPOS is a sort of Monte-Carlo (MC) generator valid for heavy ion interactions and cosmic ray air shower simulations \cite{Pierog2013}. EPOS is confronted to Relativistic  Heavy Ion Collider (RHIC) and Large Hadron Collider (LHC) data \cite{Werner2006,Pierog2013}. 

Such (MC) models are essential to analyze the acceptance of the detector, the hadrons in the universe, and other impacted effect in astrophysics all of them are confronted with high energy experiments \cite{Pierog2013}. In order to reproduce the LHC data \cite{CMS_Coll}, \cite{ALICE_coll}, \cite{ATLAS_coll} for p-p, p-Pb and Pb-Pb interactions,  Pierog et al., \cite{Pierog2013} made the necessary modification in the model. There is another promising work \cite{Dembinsk2019} for future analyzing the data from proton-oxygen (p-O) reaction at LHC energies. However in the latter they simulated the pseudo rapidity spectra of charged pions, charged kaons, and protons  at 13 TeV in p-p and p-O collisions at 10 TeV  with CRMC. 

In the present work, we utilize two different event generator EPOS $1.99$ and EPOSlhc \cite{Hanafy19} at energies ranging from $0.001$ up to $13$ TeV for 100,000 events per energy to calculate the particle ratios $k^{+} / \pi^{+}$, $k^{-} / \pi^{-}$, $\pi^{-} / \pi^{+}$, $k^{-} / k^{+}$, $\bar{p}/\pi^{-}$, $\Lambda / \pi^{-}$, $\overline{\Lambda} / \Lambda$, $\overline{\Omega} / \Omega$, $\overline{\Xi^{+}} / \Xi^{-}$, and strange and multi-strange particles such as $\phi(1020) / k_{s}^{0*}(892)$, $\overline{\Sigma}^{*+8} / \Sigma^{*+8}$, $\overline{\Sigma}^{*08} / \Sigma^{*08}$, $\overline{\Sigma}^{*-8} / \Sigma^{*-8}$, $\overline{\Xi}^{*08} / \Xi^{*08}$ and $ \overline{\Xi}^{*-8} / \Xi^{*-8}$. EPOS $1.99$ is performed at 7.7, 11.5, 19.6, 27, 39, 62.4, 130 and 200 GeV while EPOSlhc is executed at 0.9, 2.76, 5.02, 7 and 13 TeV for Pb-Pb collision. The resulting particle ratios are used to explain the strangeness enhancement signature.

\section{Results and discussion}\label{results_dis}
In this section, the obtained results of different particles ratios using the HRG model are presented from $\sqrt{s}\sim 0.001$ up to $\sqrt{s} = $ 13 TeV. All results are compared with the available experimental data. For some suggested strange and multi-strange particles, there are a lack of experimental data, thus we used two different generators, i.e., EPOS $1.99$ and EPOSlhc, to predict their results. Also, the freeze-out parameters, i.e., $T_{ch}$ and $\mu_{B}$, are estimated as a result of fitting the obtained results from the HRG model of a combination of the calculated particles ratios with both LHC data and EPOSlhc event generator results for two different collisions systems, i.e., p-p and Pb-Pb, at $\sqrt{s} =$ $5.02$, $13$ TeV, respectively. The obtained values of $T_{ch}$ and $\mu_{B}$ is compared to the values presented in Ref.\cite{Rathexp_fit}. The calculated particles ratios as a function of various center of mass energies are then used to explain the strangeness enhancement signature.

The first experimental data of strangeness enhancement in high multiplicity pp collision is presented in \cite{Adam2017} for strange and multi-strange particles. This kick-off results motivated the authors of the current work to study the strange and multi-strange particles enhancement. Additionally, they observed that there is a similarity in the strangeness production between $p-p$ and $Pb-Pb$ collisions for high multiplicity events where the deconfined phase of matter (i.e., QGP is formed). This conclusion is impacted again in different interesting works \cite{Rathexp_fit, Sharma2019}. The results are divided into three groups:

\begin{itemize}

\item Particle multiplicity versus the center of mass energy $\sqrt{s}$ \\   
Strangeness enhancement is considered as a signal of deconfinement in the ultra-relativistic heavy-ion collisions where there is an enhancement of the yields of hyperons relative to that of p-p nucleus collisions \cite{multi, shine}.
In this section, the EPOS $1.99$ event-generator is used to predict the mean multiplicities of the strange particles, $k^{+}$, $k^{-}$ and $\Lambda$, from p-p collisions at energies ranging from  $\sqrt{s}$ = $4$ up to $\sqrt{s}$ = $26$ GeV in a rapidity range of $0$ $<$ $y$ $<$ $3$ as shown in Fig. (\ref{fig:multiplicity}). The obtained results are confronted to that measured in NA61/SHINE experiment \cite{multi, shine}. EPOS $1.99$ event-generator is succeeded very well to describe the multiplicity of $k^{+}$ as seen in Fig. (\ref{fig:multiplicity}a). In case of $k^{-}$, there is a small deviation at $\sqrt{s}$ $=$ $9$ and $19$ GeV as shown in Fig. (\ref{fig:multiplicity}b). The multiplicity of $\Lambda$ particle predicted by EPOS $1.99$ event-generator is shown in Fig. (\ref{fig:multiplicity}c) and has a good agreement with the experimental data taken from \cite{multi, shine}.

\begin{figure}[htbp]
\includegraphics[width=8.cm]{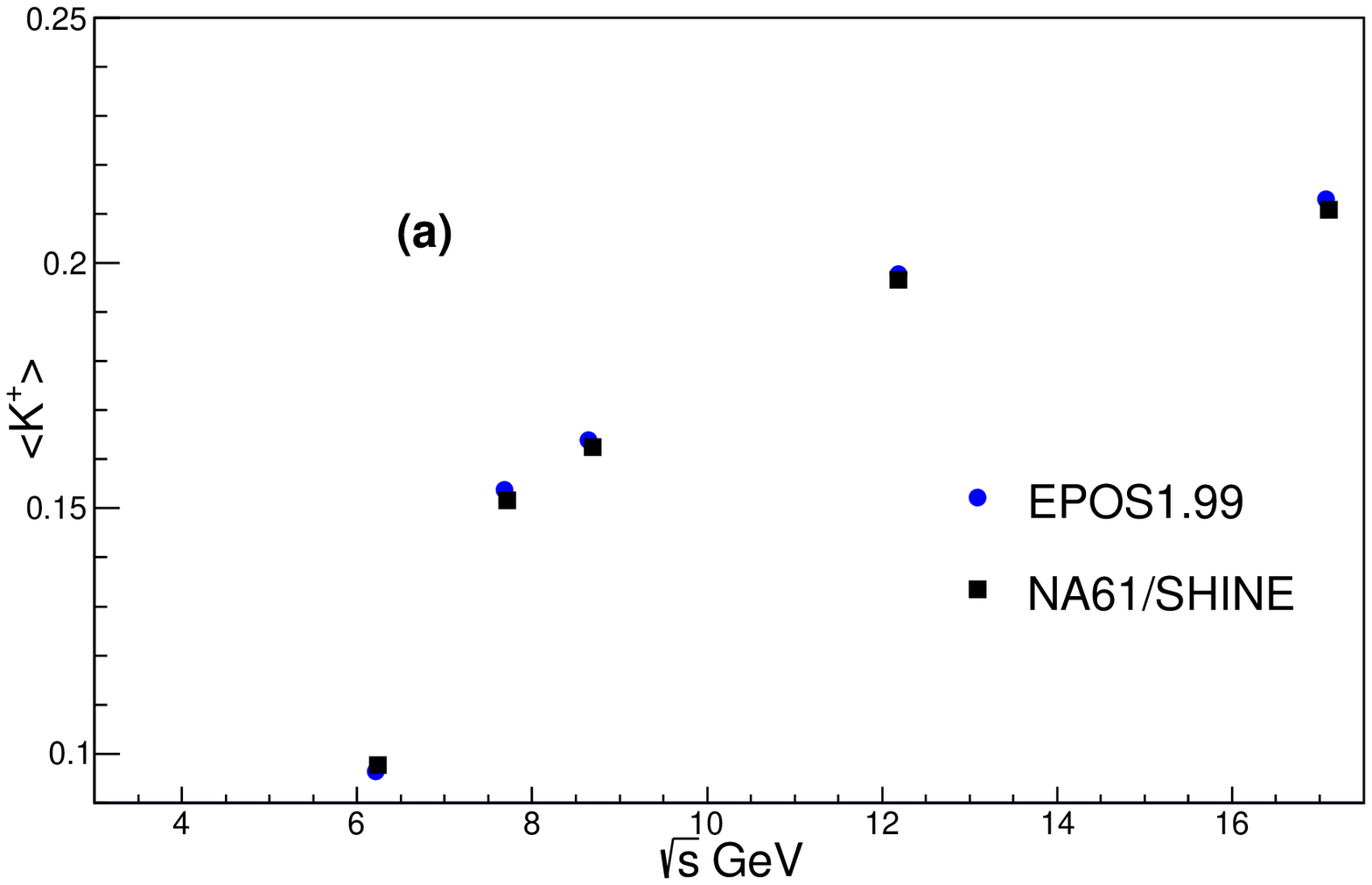}
\includegraphics[width=8.cm]{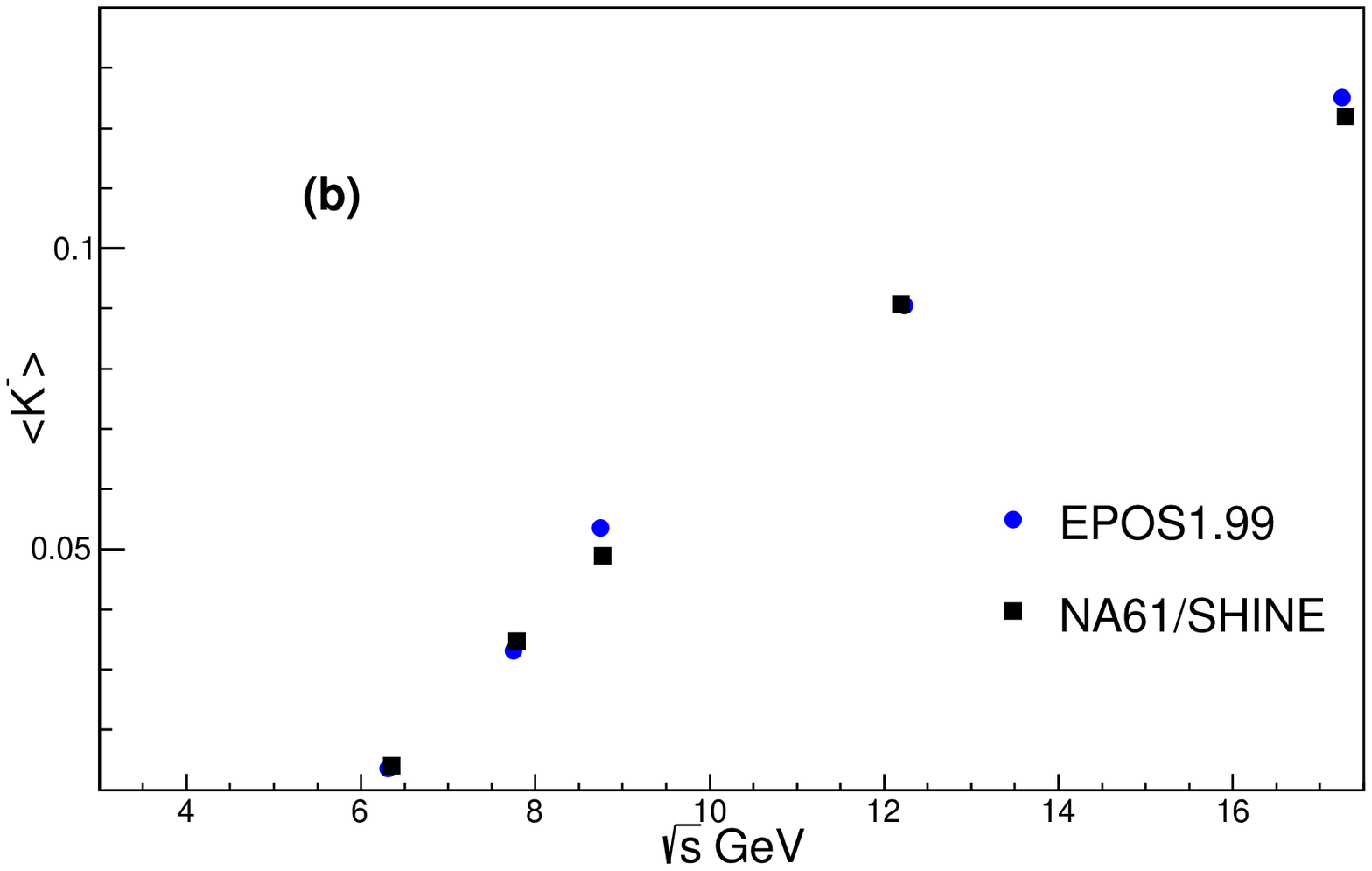}
\center \includegraphics[width=8.cm]{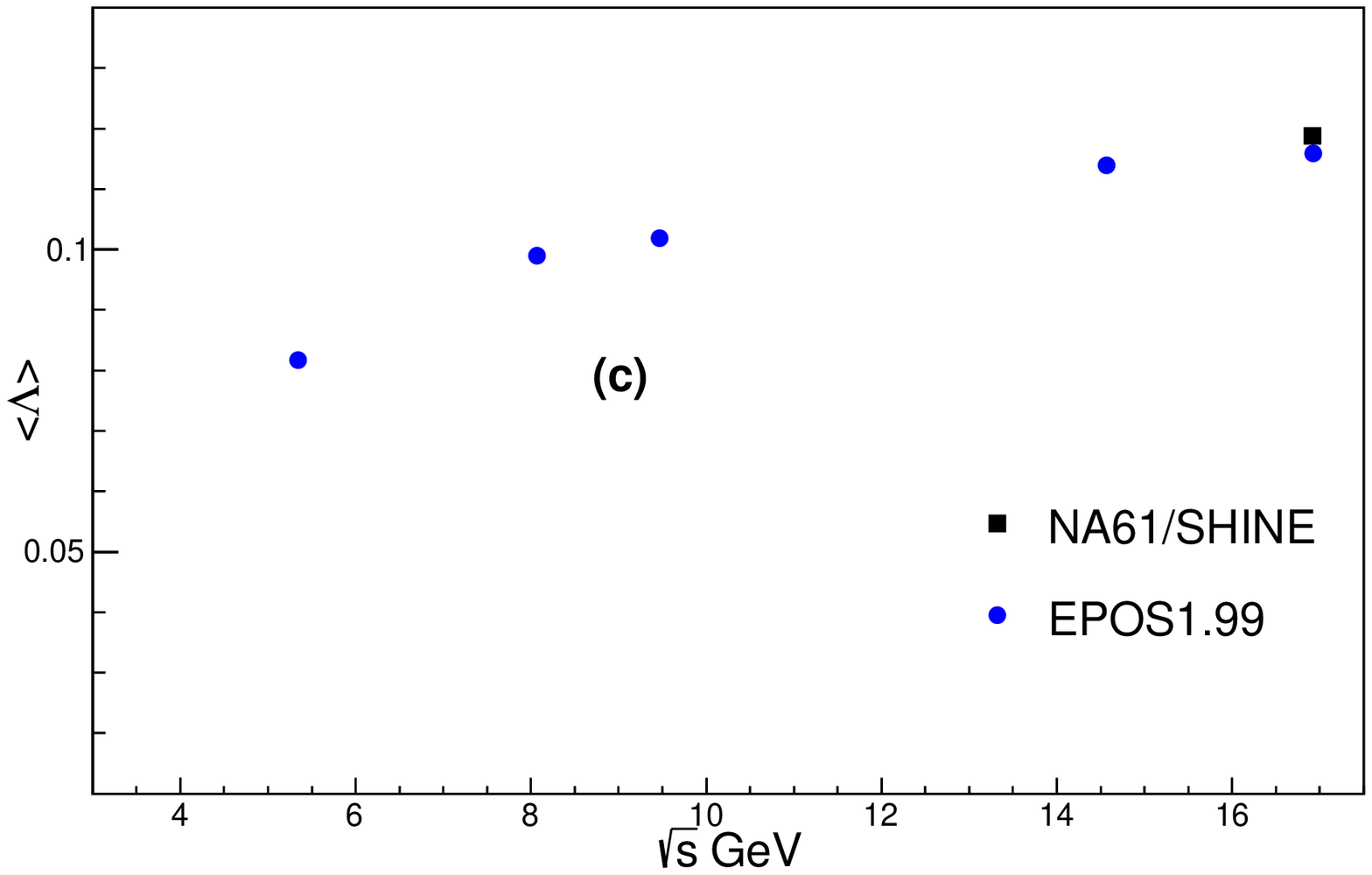}
\caption{The mean multiplicity of the particles (a) $k^{+}$, (b) $k^{-}$ and (c) $\Lambda$ generated from EPOS $1.99$ in comparison with the experimental data taken from \cite{multi, shine}.}
\label{fig:multiplicity}
\end{figure}

\end{itemize}

\begin{itemize}
\item Particle ratios versus the center of mass energy $\sqrt{s}$ \\
 The particle ratios with the $\sqrt{s}$ included some heavier and strange particles are calculated by the HRG model and both event generators, i.e., EPOS $1.99$ and EPOSlhc, at various energies spanning from  $\sqrt{s}$ = $0.001$ up to $\sqrt{s}$ = $13$ TeV. 
 The dependence of the baryon chemical potential and the temperature on the center of mass energy is taken from \cite{Tawfik15}, which has an agreement with the paramterization in \cite{cly}
 \begin{equation}
 \mu_{B}= \frac{a}{1+b \sqrt{s}}
 \end{equation}
 Where $a=1.245 \pm 0.094$ GeV, and $b=0.264 \pm 0.028$ $GeV^{-1}$. The temperature can also be defined in terms of the center of mass energy \cite{Tawfik15}.
 \begin{equation}
 T = T_{lim} \left[ \frac{1}{1+exp\left( \frac{1.172-log(\sqrt{S_{NN}}}{0.45})\right) }\right] 
 \end{equation}
  where $\sqrt{S_{NN}}$ is taken in GeV and $T_{lim} =  161\pm 4\hspace{0.05cm}MeV$. 
The quark structure of the strange and multi-strange particles suggested here and listed in Tab. (\ref{Table1}).

 \begin{table}[htbp]
 \begin{center}
\caption{The strange and multi-strange particles with their quark structure and masses.}
    \label{Table1}
 \begin{tabular}{|c|c|c|} 
 \hline
 Particle & Quark Content  & mass [GeV]\\  
 \hline
 $\phi(1020)$ & $ s\overline{s} $ & 1.01946\\ 
 \hline
 $  k_{s}^{0*}(892)$ & $u\overline{s}  $& 0.89594\\
 \hline
 $ \Lambda $ & $dus$& 1.11568\\
 \hline
 $ \Omega $ &  $sss$&1.67243\\
 \hline
 $\Xi^{-}$ &$ dss$ & 1.32171\\
 \hline
 $\Sigma^{*+8}$ & $  uus$ &1.3828\\
 \hline
 $\Sigma^{*08}$ & $dus  $ &1.3837\\
 \hline
 $\Sigma^{*-8}$ & $dss  $ &1.3872 \\
 \hline
 $\Xi^{*08}$ & $ uss $ & 1.5318\\ 
 \hline
  $\Xi^{*-8}$ & $ dss $ & 1.535	\\
 \hline
\end{tabular}
\end{center}
\end{table}

The dependence of different particle ratios on $\sqrt{s}$ at LHC energies from GeV to TeV is studied utlizing the HRG model. Fig. (\ref{fig:kaon_pion}) illustrates the ratios of strange to non-strange particles (upper panel) such as $K^{+}/\pi^{+}, \hspace{0.1cm} K^{-}/\pi^{-}$ and pure non-strange and strange ratios (lower panel) for $\pi^{-}/\pi^{+}, \hspace{0.1cm} K^{-}/K^{+} $ versus the center of mass energy. These ratios are confronted to the experimental data \cite{exp1,exp2,exp3,exp4}, and EPOS $1.99$ (used at low energies) and EPOSlhc (used at high energies) event generators from $ 10^{-3}\leqslant \sqrt{s} \leqslant $ 13 TeV. Fig. (\ref{fig:kaon_pion} a) shows the important particle ratio of $K^{+}/\pi^{+}$ which is used as characterising tool to describe the strangeness enhancement in the quantum chromodynamic (QCD) matter \cite{multi}. This ratio shows a peak at $\sqrt{s}$ $\simeq$ 9 GeV which is known as the horn puzzle and might be considered as an indication of the QCD phase transition \cite{multi}. The EPOS $1.99$ event-generator can describe the lowest NA61/SHINE data produced from p-p collisions at centre of mass energy $\sqrt{s}$ $\simeq$ $7.5$ and $12.5$ GeV, and the highest STAR Pb-Pb collisions at $\sqrt{s}$ $\simeq$ $200$ GeV while the EPOSlhc event-generator can describe the ALICE data at $\sqrt{s}$ $\simeq$ $3$ TeV. The wide range of energy shows the expected results in which there is a rapid enhancement in the strange particles only in the ratios ($K^{-}/\pi^{-}, K^{-}/K^{+} $) as in fig. (\ref{fig:kaon_pion}b,d). However, Fig. (\ref{fig:kaon_pion} a) shows a monotonic increasing (horn) up to $\sqrt{s}\sim $ 10 \hspace{0.2cm} GeV, then begin to decrease with increasing the energy, and a clear deceasing in the pure non-strange particles with increasing the energy such as fig.(\ref{fig:kaon_pion}c).

\begin{figure}[htbp]
\includegraphics[width=8.cm, height=5.cm]{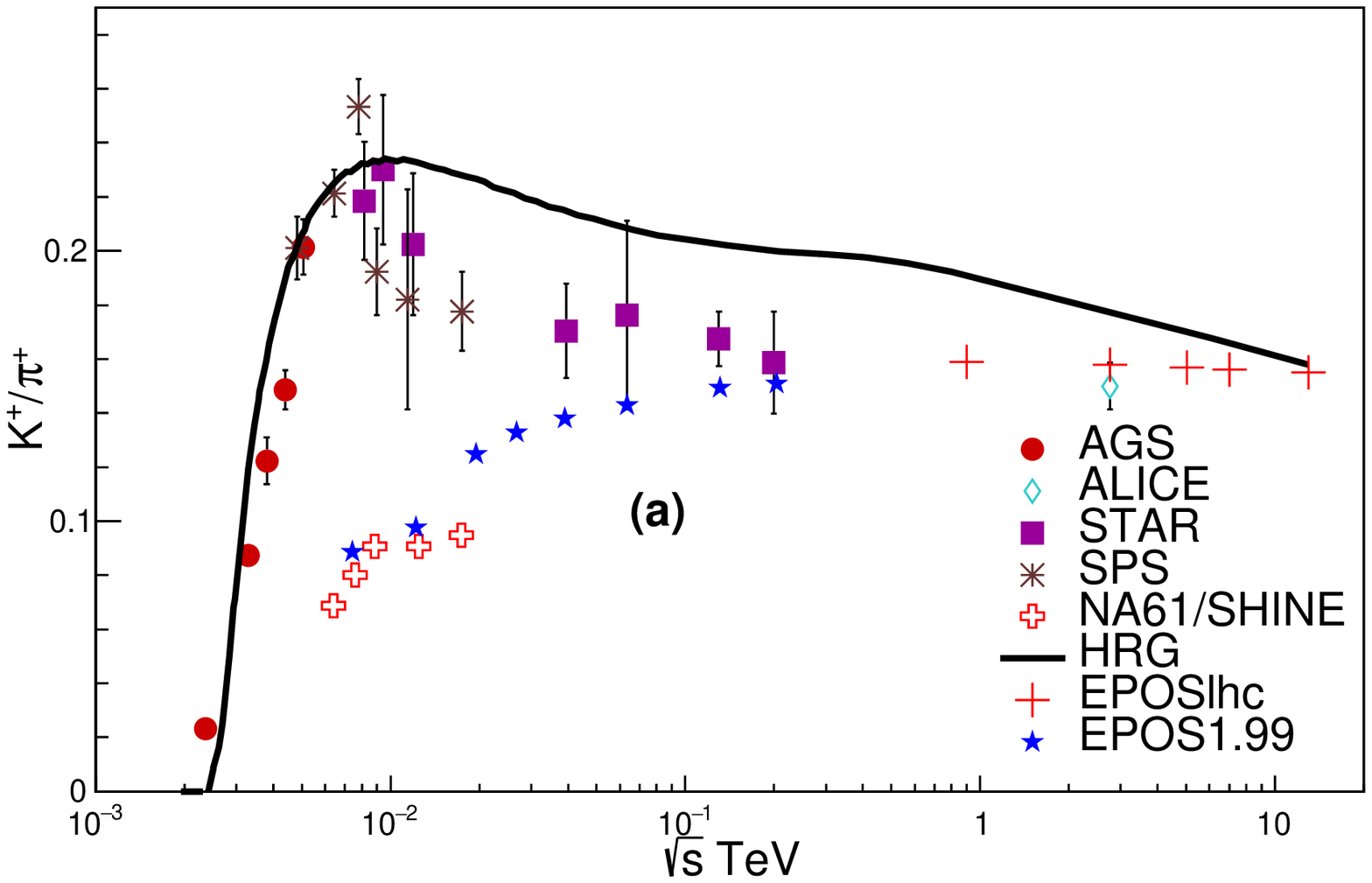}
\includegraphics[width=8.cm]{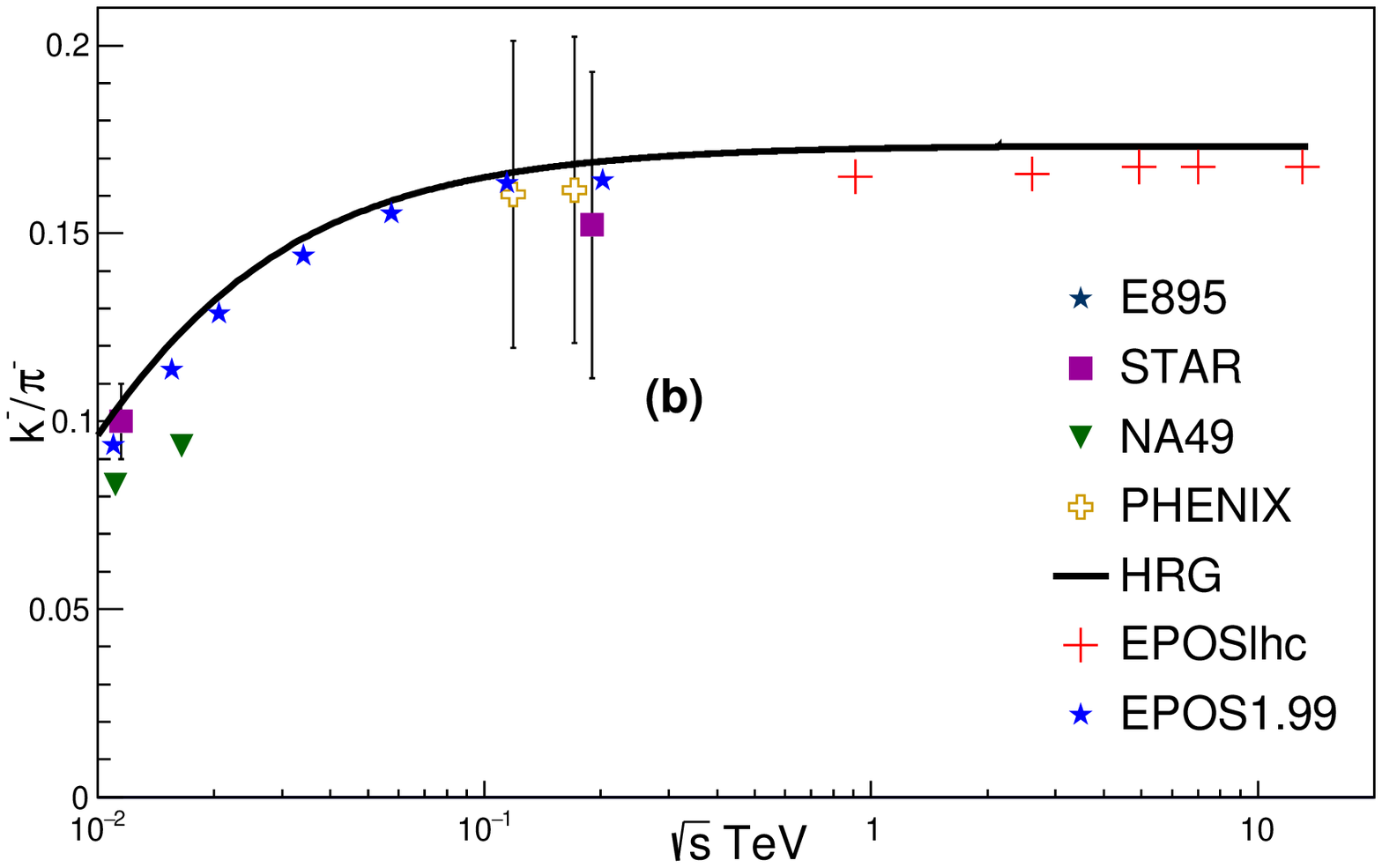}
\includegraphics[width=8.cm]{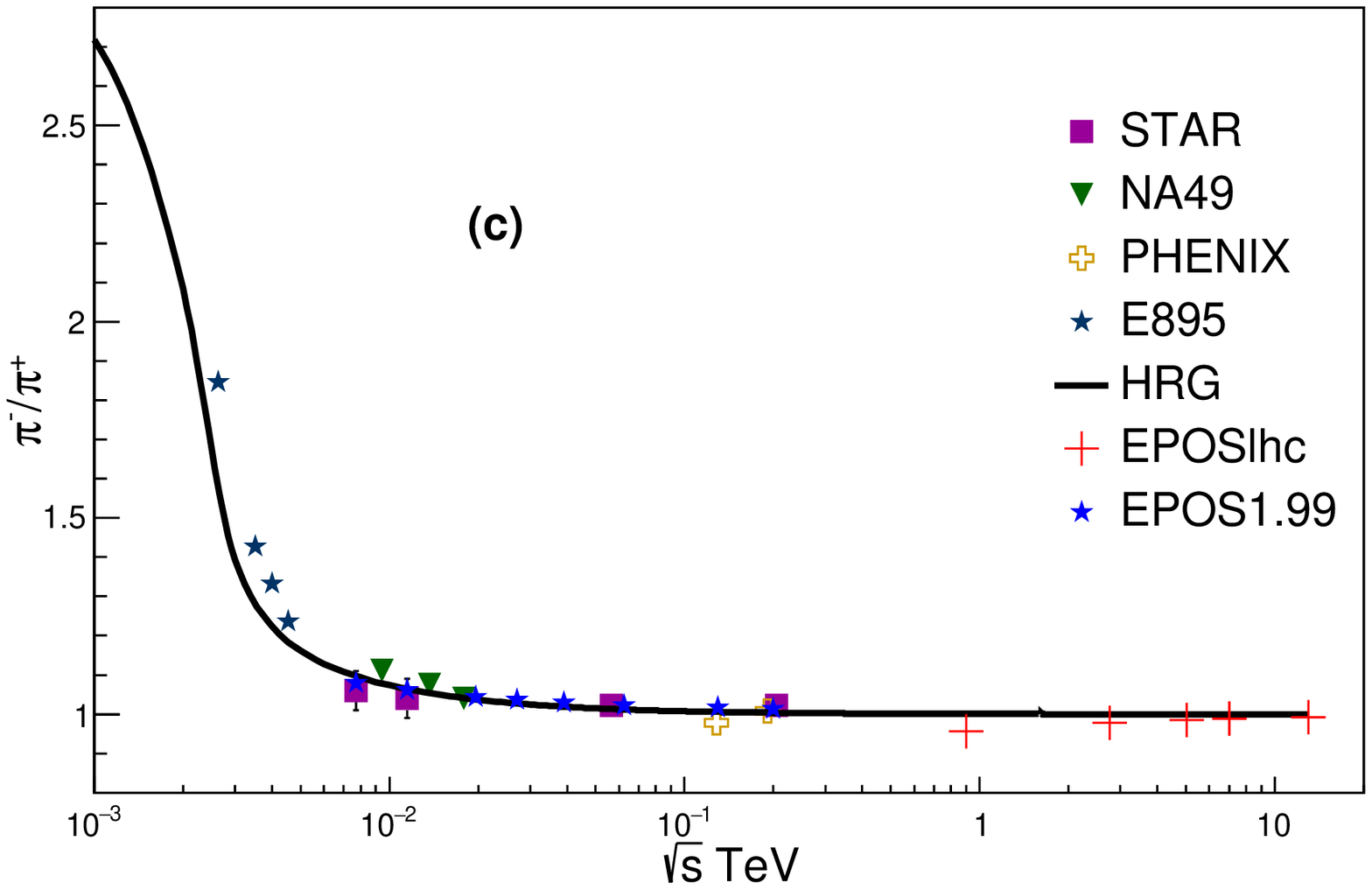}
\includegraphics[width=8.cm]{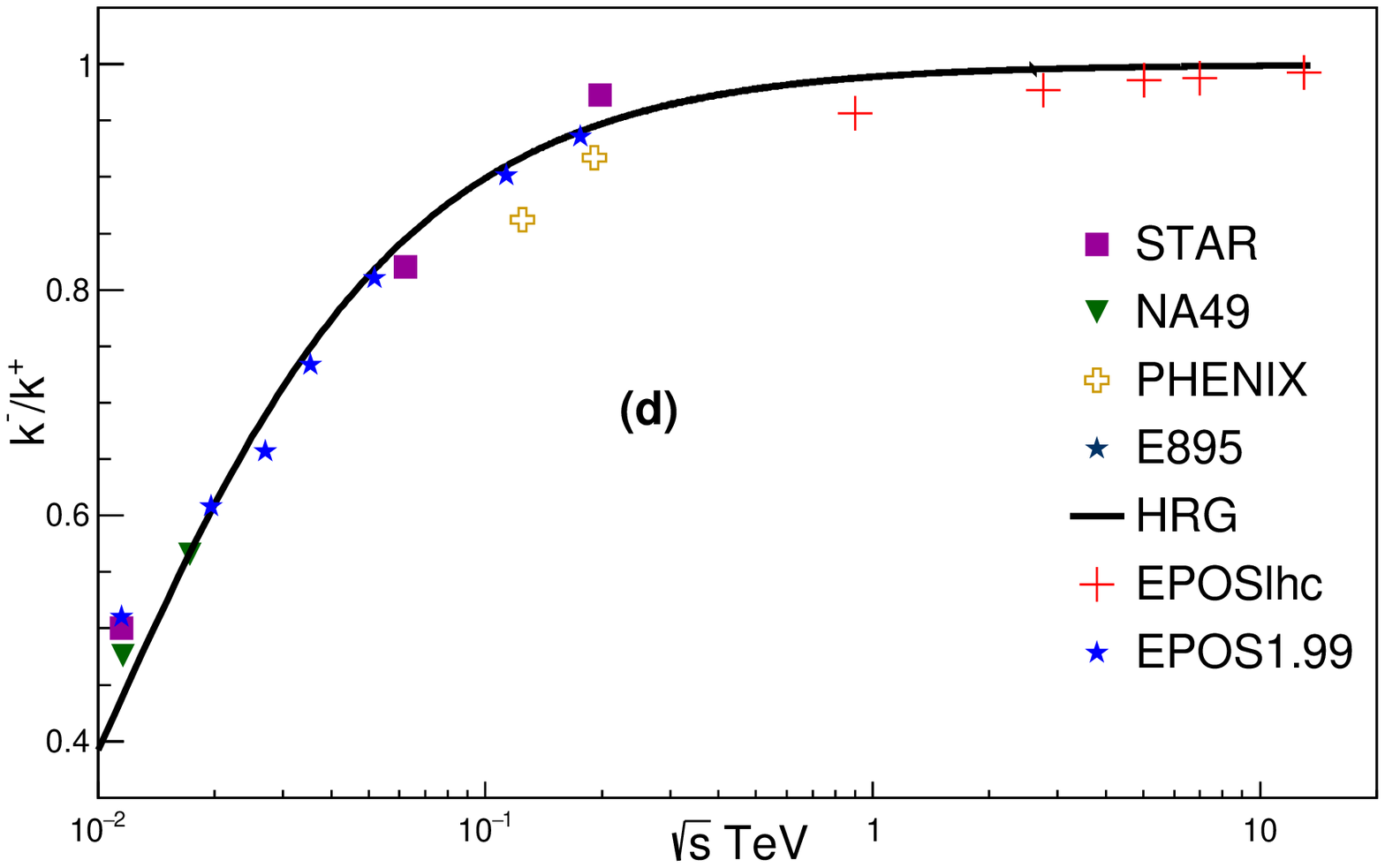}
\caption{The energy dependence of the particle ratios $k^{+} / \pi^{+}$, $k^{-} / \pi^{-}$, $\pi^{-} / \pi^{+}$ and $k^{-} / k^{+}$ from HRG model, and EPOS $1.99$ and EPOSlhc event-generators in comparison with the experimental data taken from \cite{exp1,exp2,exp3,exp4}.}
\label{fig:kaon_pion}
\end{figure}

Fig. (\ref{fig:First}) presents the energy dependence of the particle ratios $\Lambda / \pi^{-}$, $\bar{p}/\pi^{-}$ and $\overline{\Lambda} / \Lambda$ in comparison with the experimental data taken from \cite{exp1,exp2,exp3,exp4} and the estimated results from both EPOS $1.99$ and EPOSlhc event-generators. We notice that the horn puzzle appears again in the ratio $\Lambda / \pi^{-}$ still at the range of GeV energy.
\begin{figure}[htbp]
\includegraphics[width=8.cm]{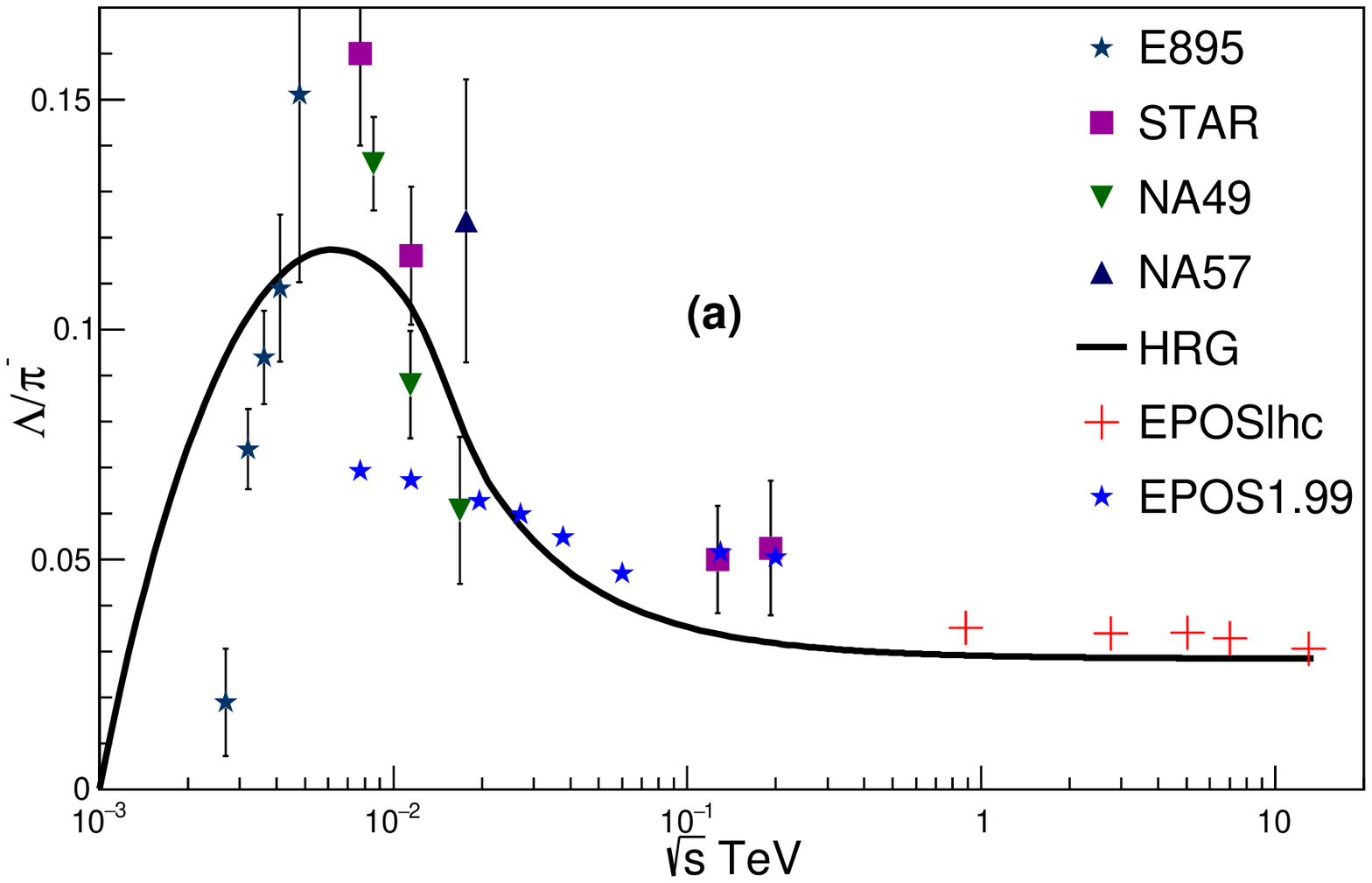}
\includegraphics[width=8.cm]{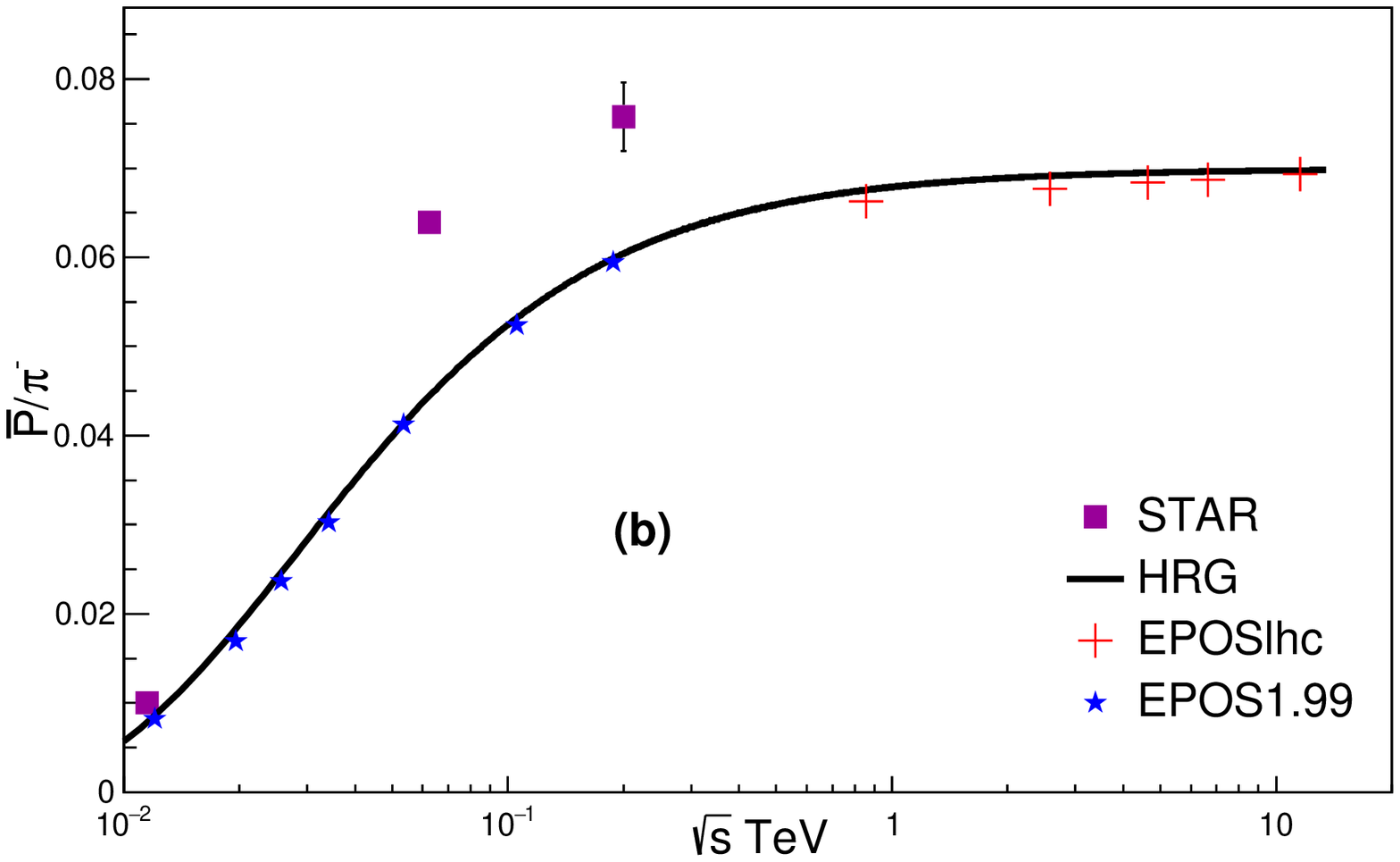}
\center{\includegraphics[width=8.cm, height=5.cm]{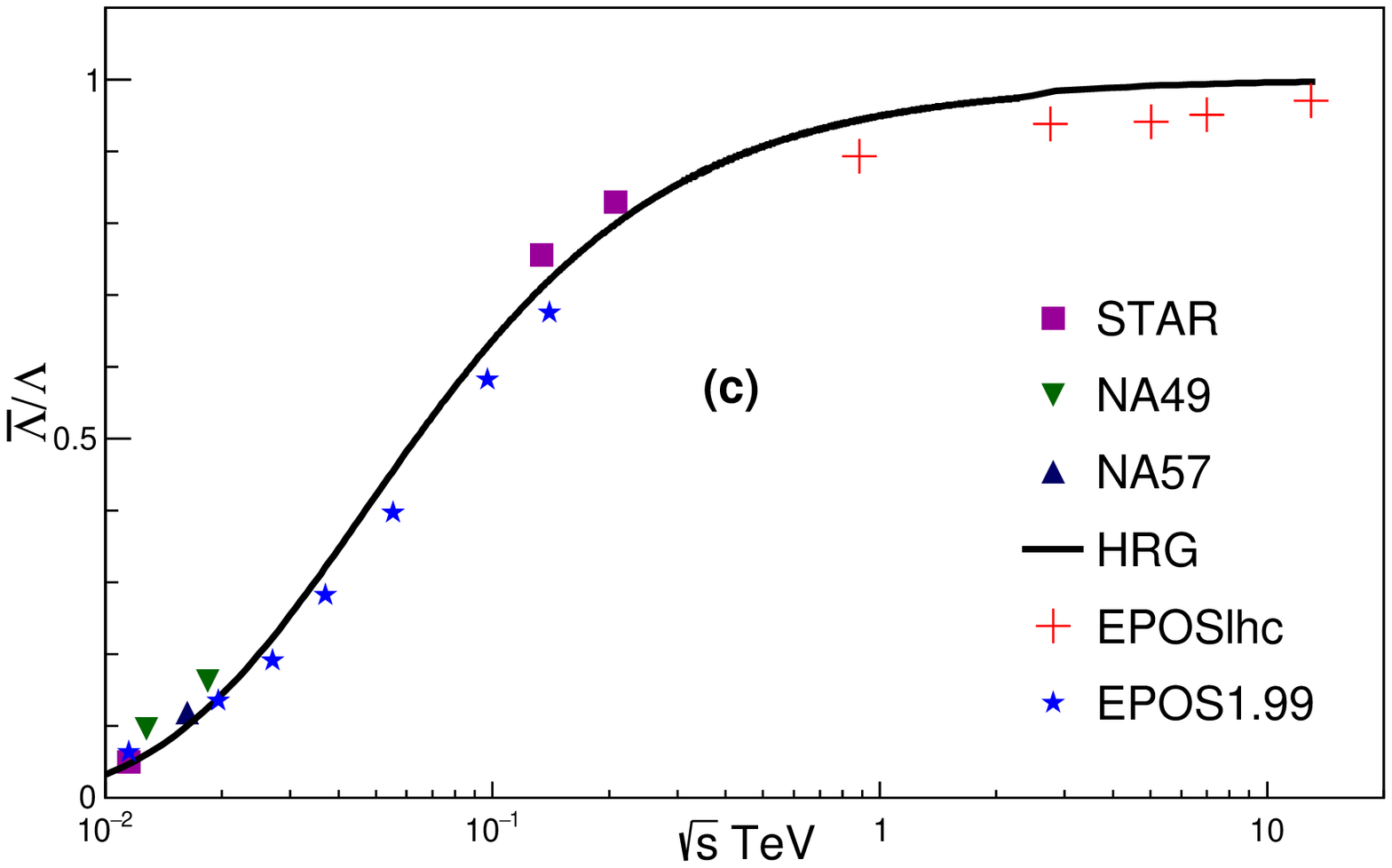}}
\caption{Similar to Fig. (\ref{fig:kaon_pion}), Particles ratios $\Lambda / \pi^{-}$, $\bar{p}/\pi^{-}$ and $\overline{\Lambda} / \Lambda$ where the experimental data are taken from \cite{exp1,exp2,exp3,exp4,exp5,exp6,exp7,exp8,exp9,exp10}.}
\label{fig:First}
\end{figure}

Figs. (\ref{fig:Second} and \ref{fig:Third}) show a series of strange and multi-strange particles such as $\overline{\Omega} / \Omega$, $\overline{\Xi^{+}} / \Xi^{-}$, $\phi/k_{s}^{*}$, $\overline{\Sigma}^{*+8} / \Sigma^{*+8}$, $\overline{\Sigma}^{*08} / \Sigma^{*08}$, $\overline{\Sigma}^{*-8} / \Sigma^{*-8}$, $\overline{\Xi}^{*08} / \Xi^{*08}$ and $ \overline{\Xi}^{*-8} / \Xi^{*-8}$ are calculated in the framework of the HRG model and compared with the results obtained form both EPOS $1.99$ and EPOSlhc event-generators. It is clear that most of the strange and multi-strange particles show strangeness enhancement as the energy increases up to 13 \hspace{0.05cm} TeV. The $\phi(1020)/ K_{s}^{0*}(892)$ \cite{ALICE2017} ratio shows a rapid enhancement at energies in  GeV and smoothly increases at TeV. This ensures that the strangeness enhancement is a strong signature for the quark gluon plasma (QGP) creation at very high energy.  
\begin{figure}[htbp]
\includegraphics[width=7.cm]{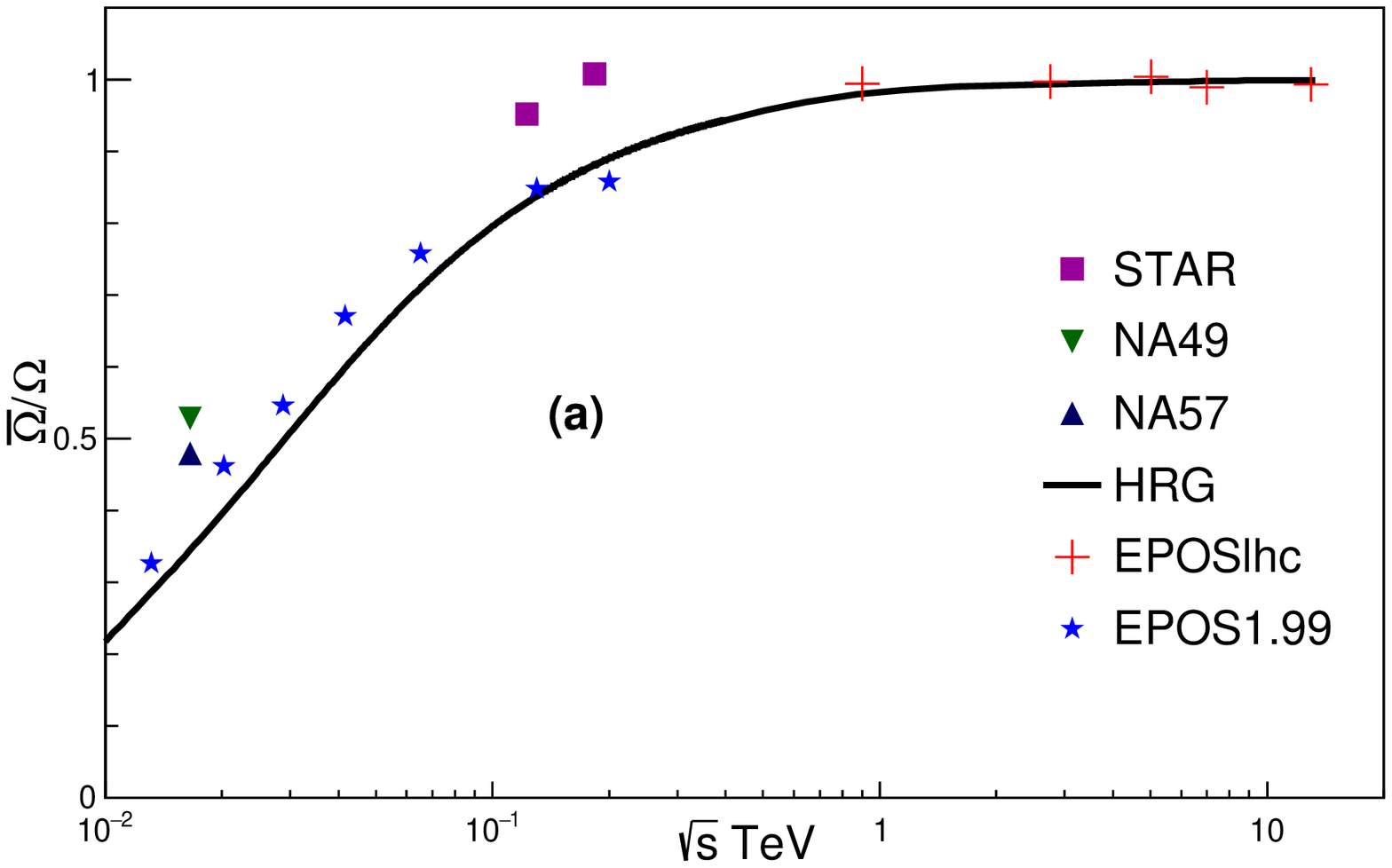}
\includegraphics[width=7.cm]{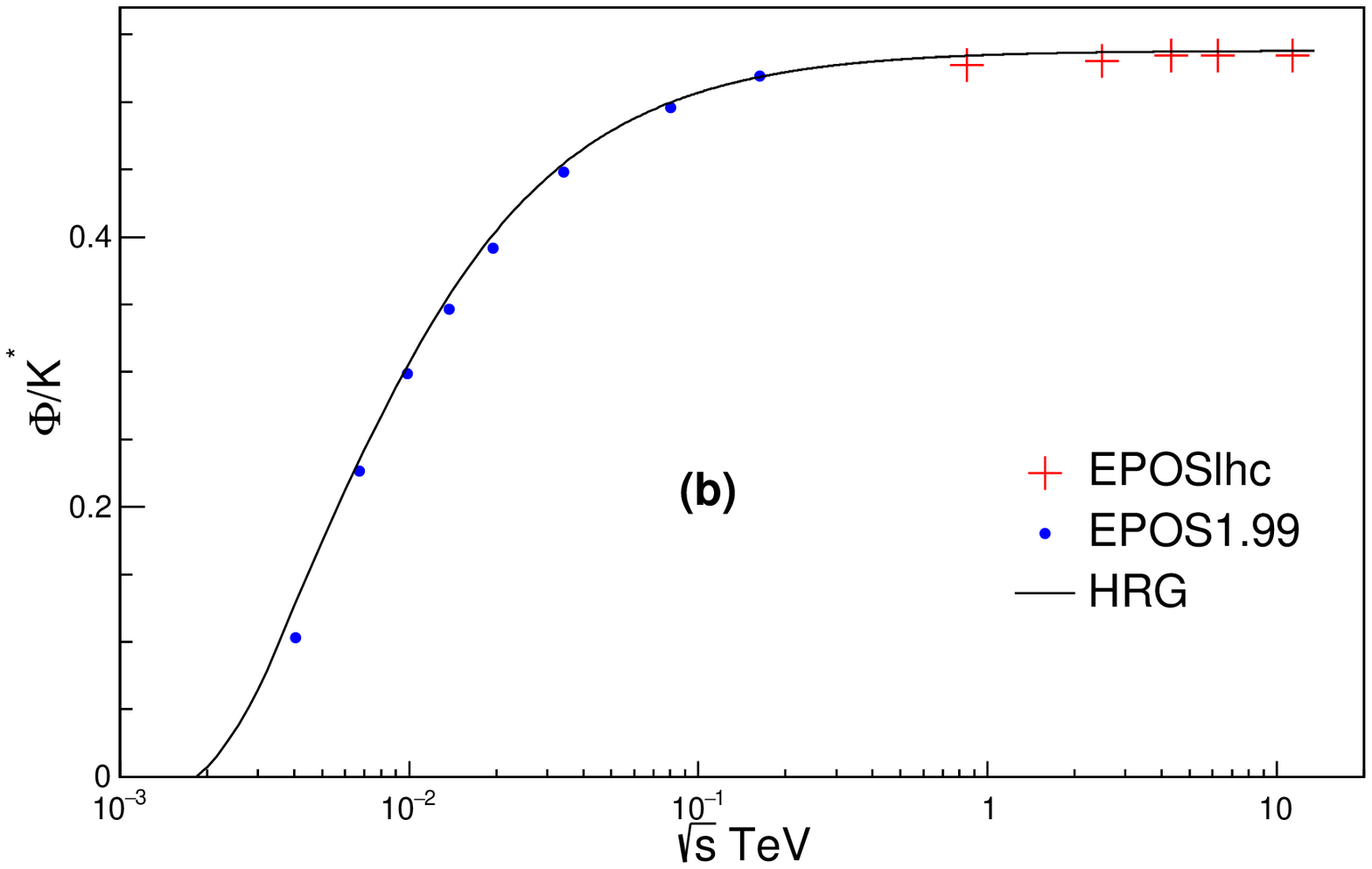}
 \centering
\includegraphics[width=7.cm]{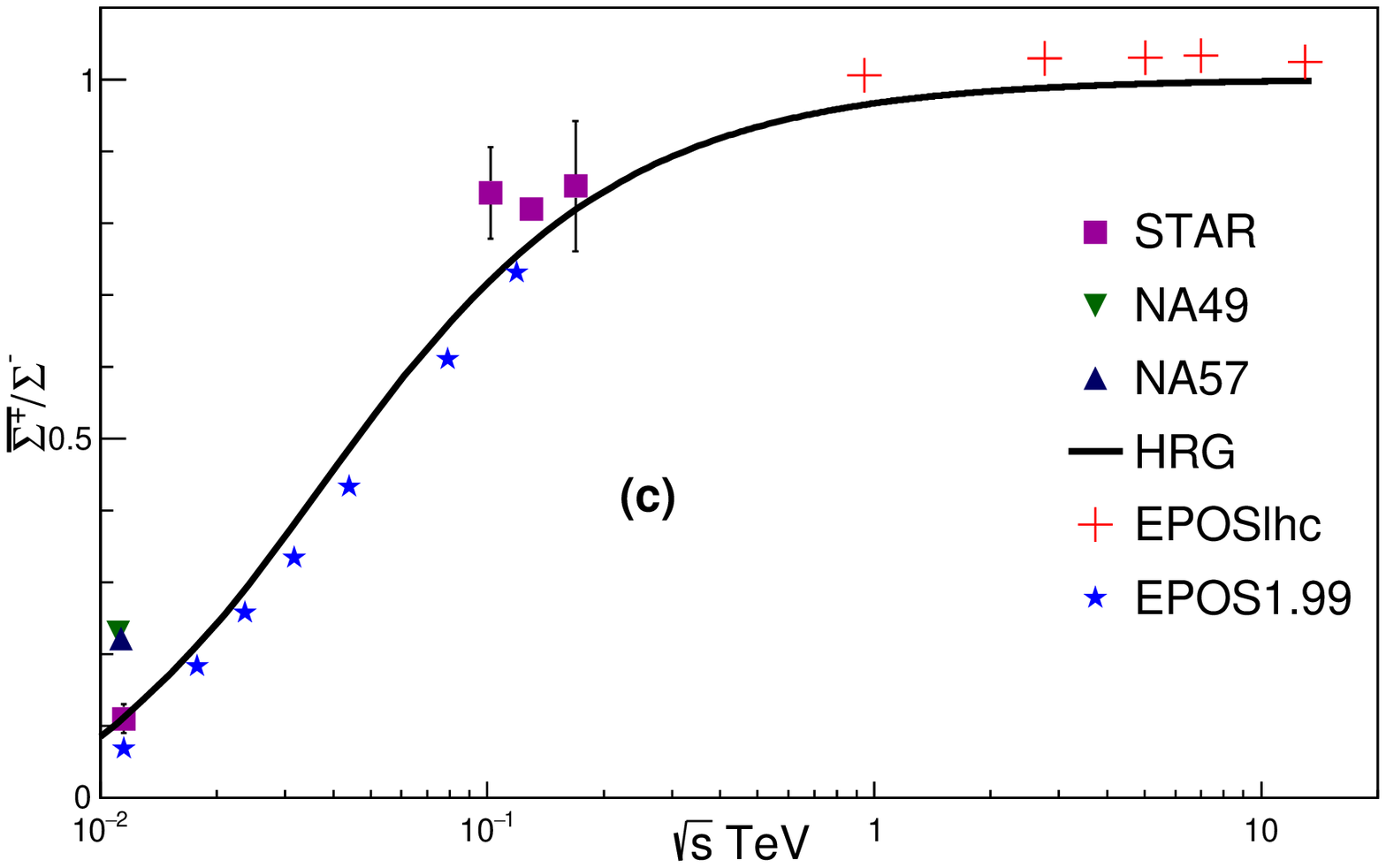}
\caption{The same as in fig. (\ref{fig:First}) the particle ratios $\overline{\Omega} / \Omega$, $\overline{\Xi^{+}} / \Xi^{-}$ and $\phi/k_{s}^{*}$ where the experimental data are taken from \cite{exp10,exp11,exp12}. In case of the particle ratio $\phi/k_{s}^{*}$, no experimental data are available and both EPOS $1.99$ and EPOSlhc are utilized for predictions.}
\label{fig:Second}
\end{figure}

\begin{figure}[htbp]
\includegraphics[width=7.cm]{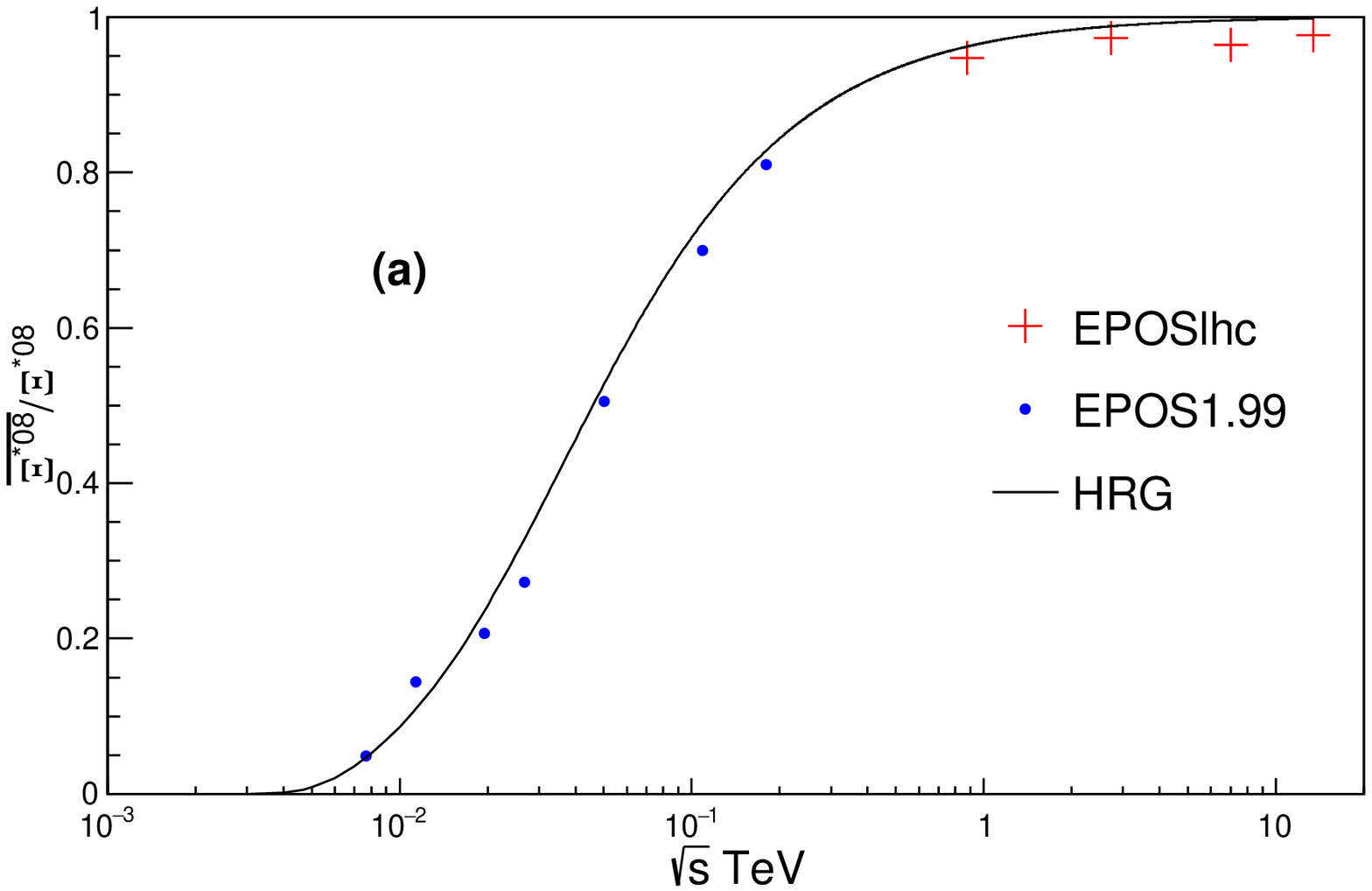}
\includegraphics[width=7.cm]{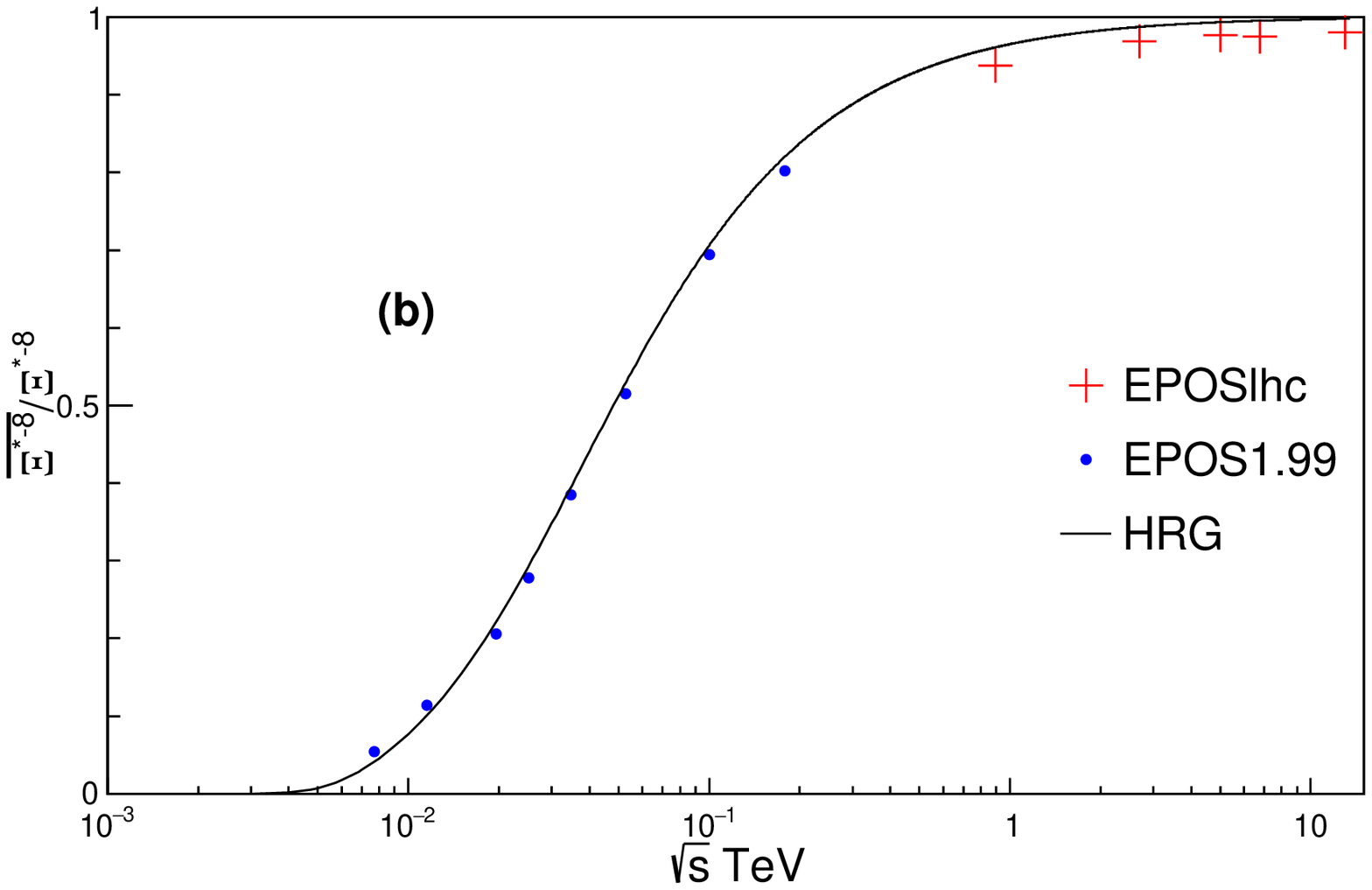}
\includegraphics[width=7.cm]{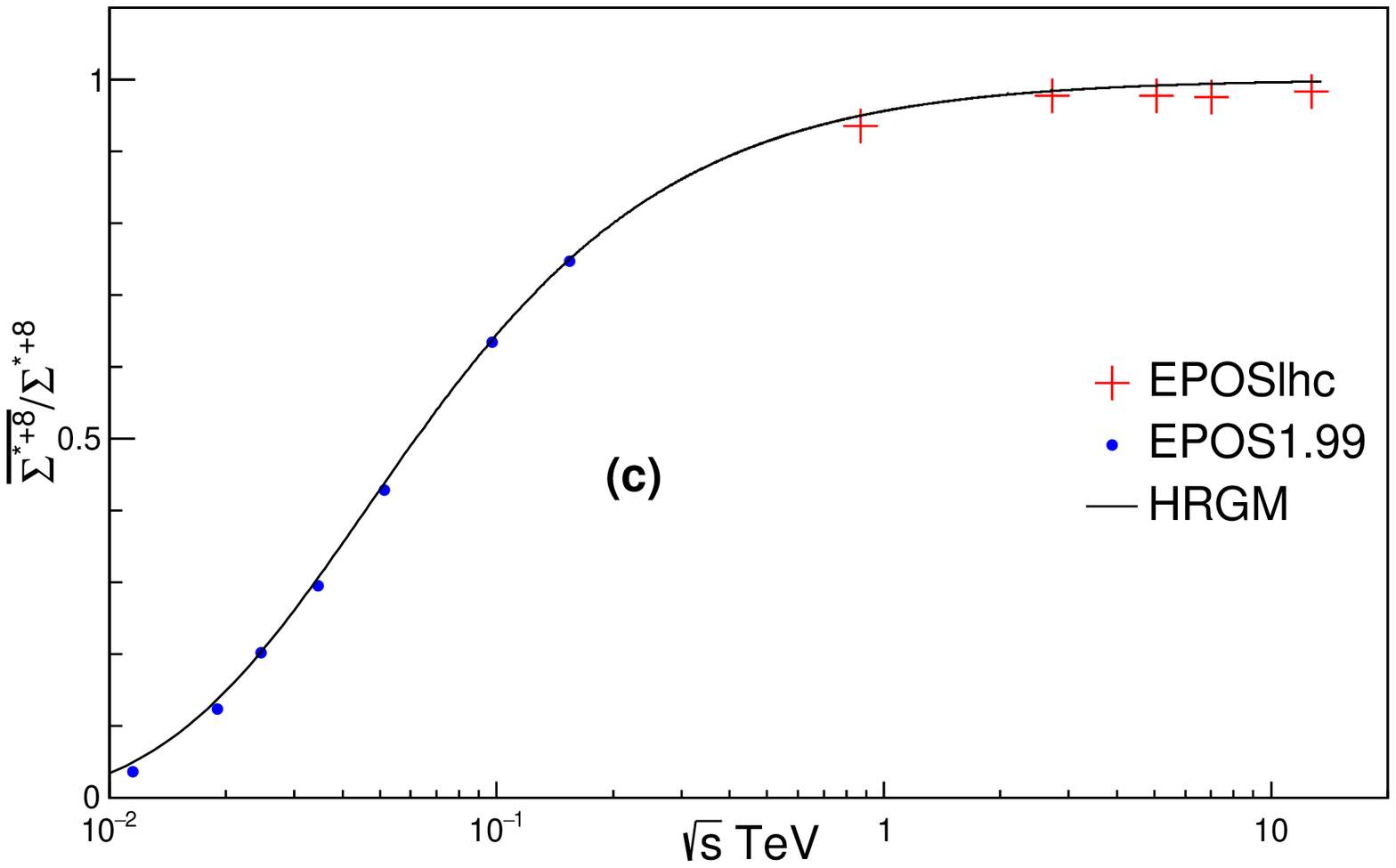}
\includegraphics[width=7.cm]{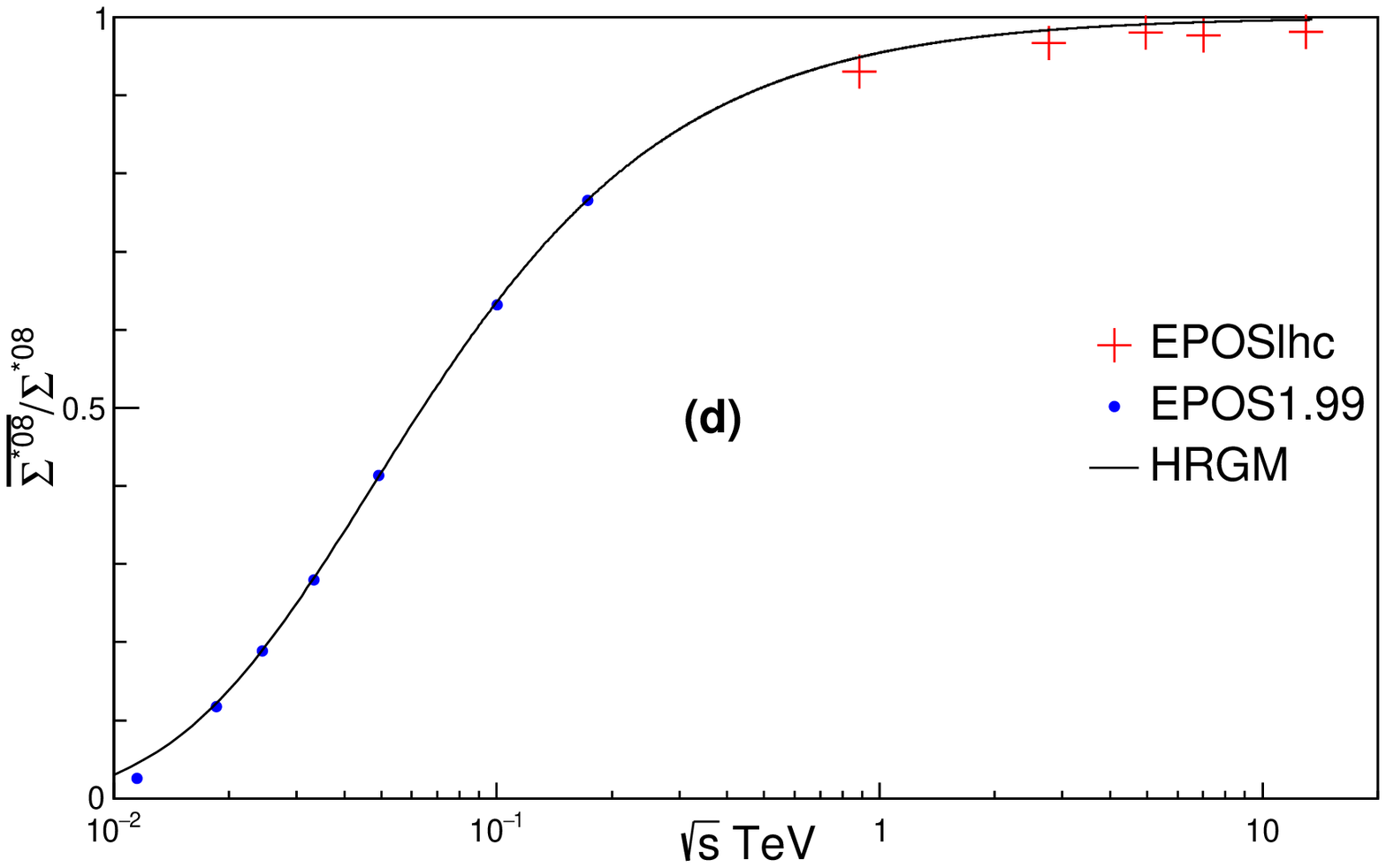}
 \centering
\includegraphics[width=7.cm]{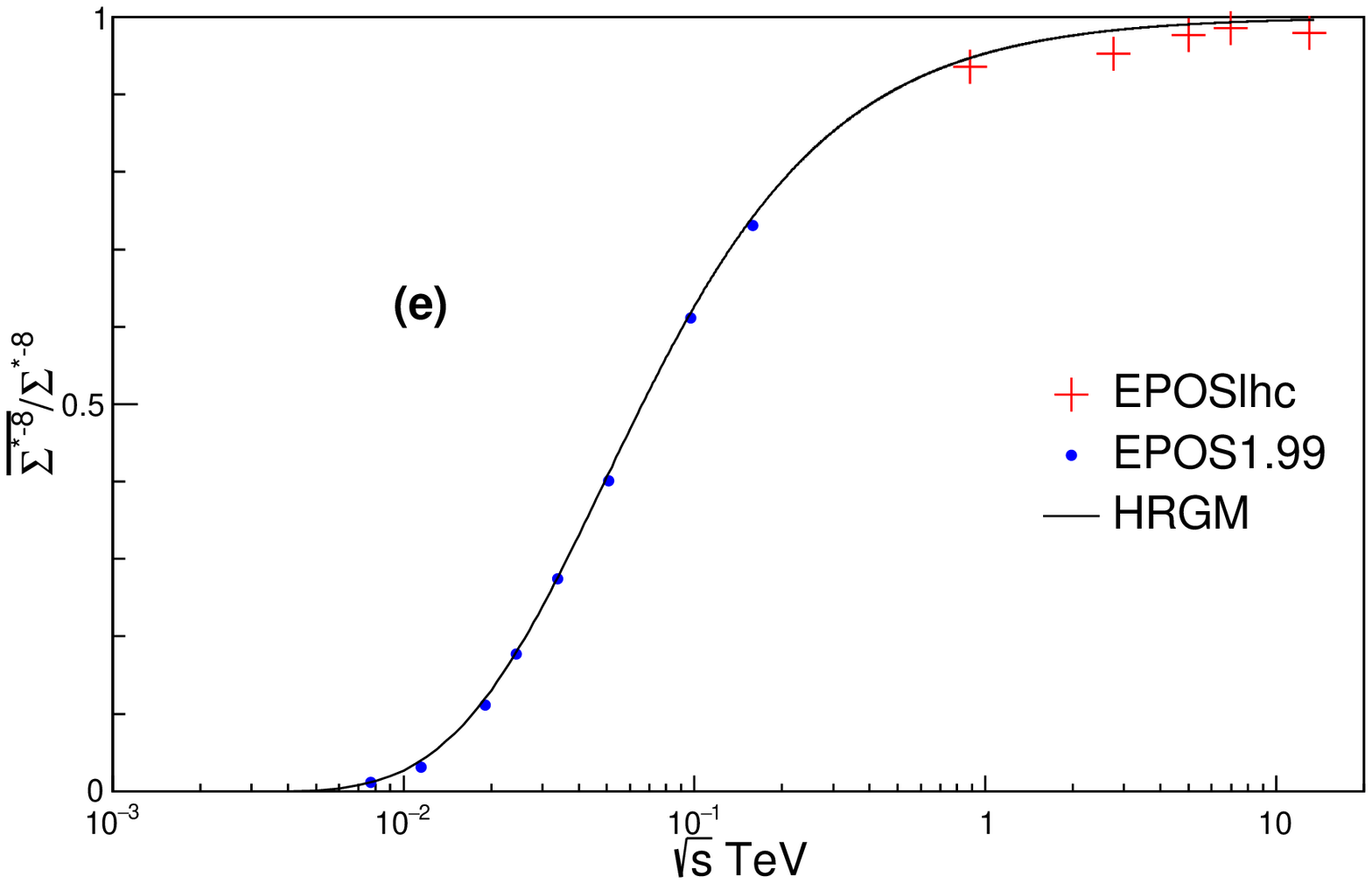}
\caption{The same as in  Fig. (\ref{fig:Second}) particle ratios $\overline{\Sigma}^{*+8} / \Sigma^{*+8}$, $\overline{\Sigma}^{*08} / \Sigma^{*08}$, $\overline{\Sigma}^{*-8} / \Sigma^{*-8}$, $\overline{\Xi}^{*08} / \Xi^{*08}$, and $ \overline{\Xi}^{*-8} / \Xi^{*-8}$.}
\label{fig:Third}
\end{figure}

\item Fitting $\chi^{2}$-tuning\\
 Recently, fitting with particle ratios for both $p-p$ and $Pb-Pb$ collisions has been made in Ref. \cite{Sharma2019, Sharma2018}. They found that the HRG model fits very well and the grand canonical description is valid for the highest multiplicities. Fig. (\ref{fig:ratios}) shows a fitting between the calculated particles ratios, i.e., particles shown in Tab. (\ref{Table2}), from the HRG model using Eq. (\ref{eq:5}), and both LHC data \cite{Rathexp_fit} and EPOSlhc event generator results, using Eq. (\ref{eq:fitttt}) in $Pb-Pb$ and $p-p$ collision systems at energies $5.02$ and $13$ TeV, respectively. 
\begin{figure}[htbp]
\includegraphics[width=8.cm]{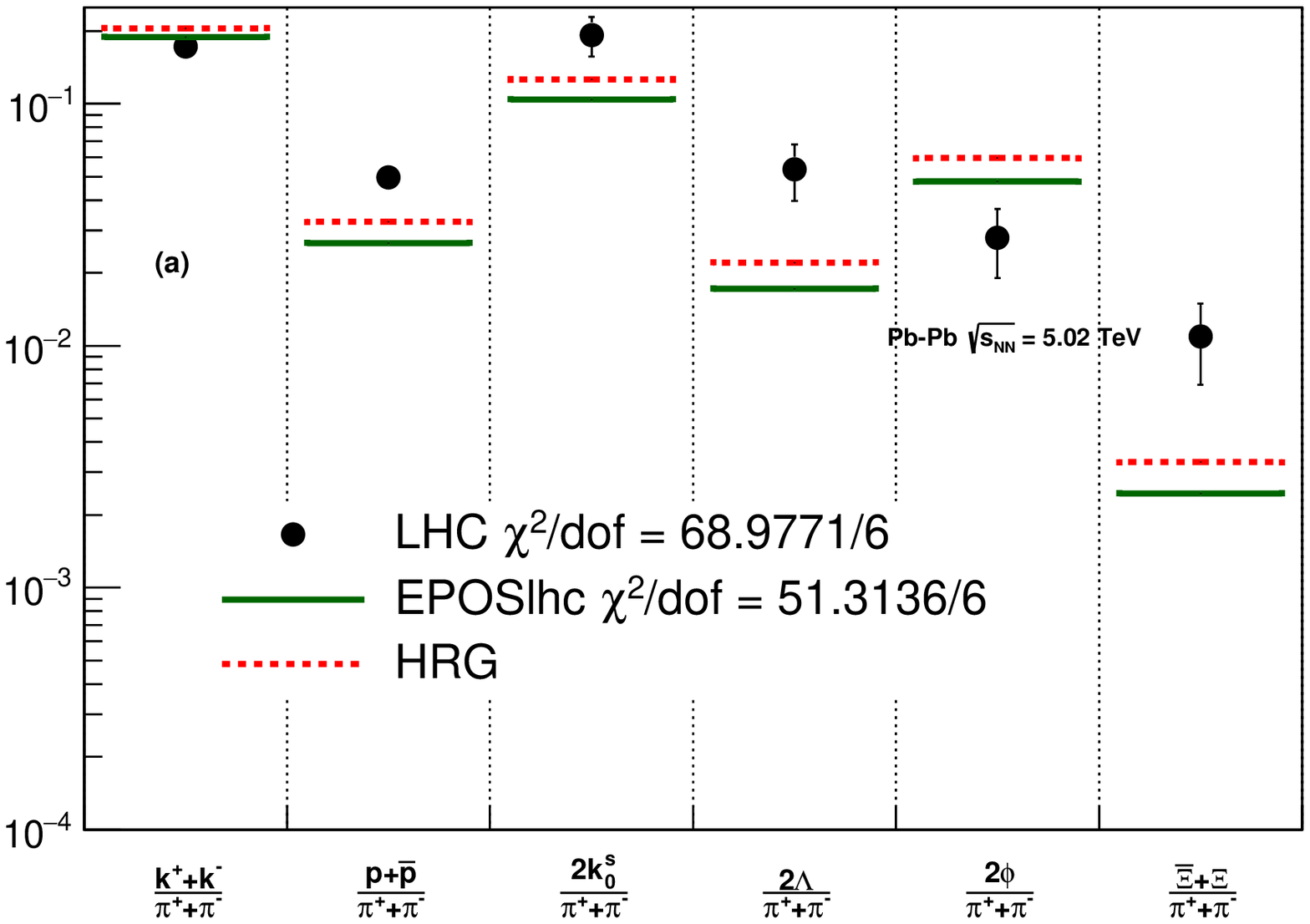}
\includegraphics[width=8.cm]{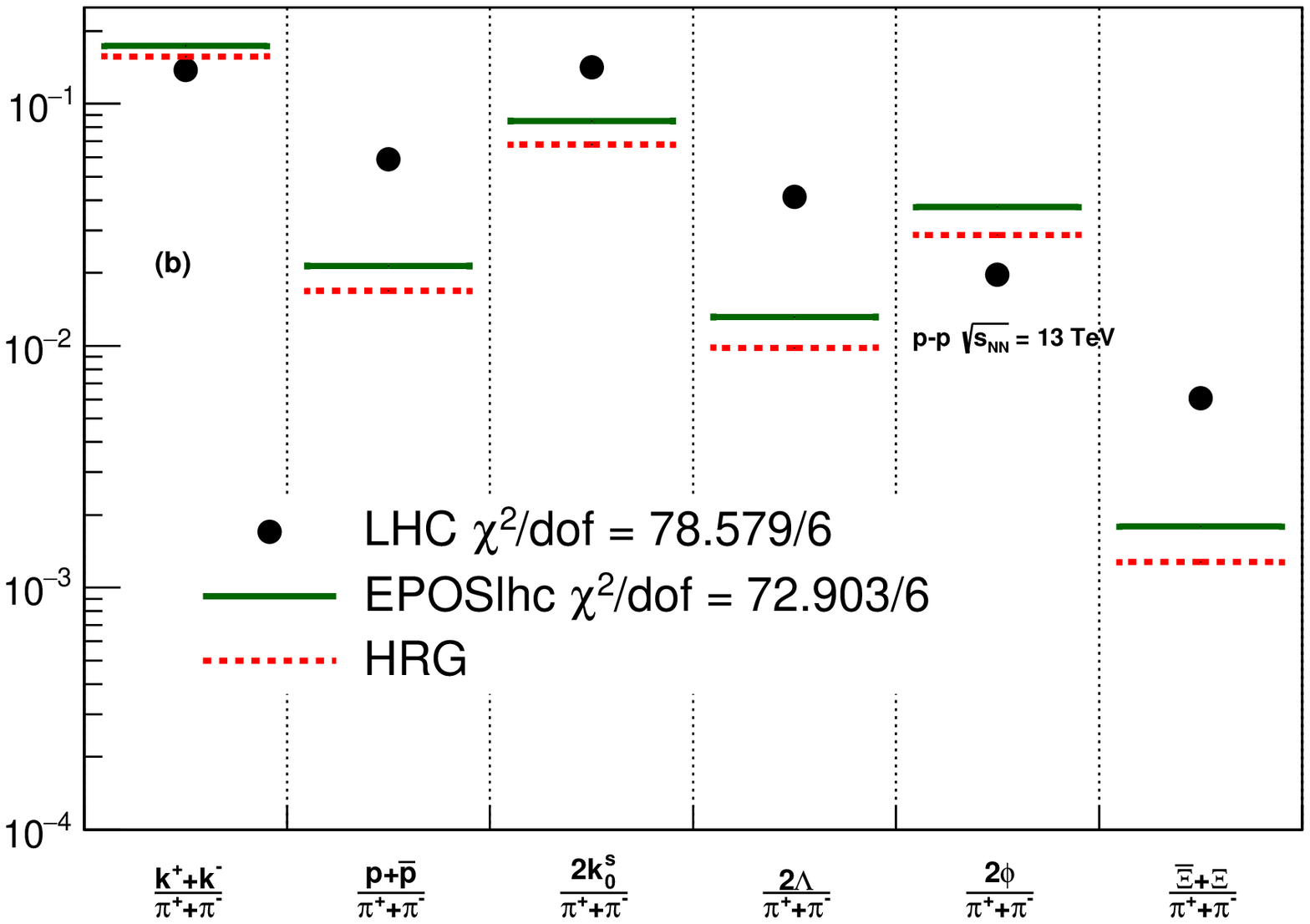}
\caption{A statistical fit between the calculated particles ratios using HRG model Eq. (\ref{eq:5}), and both LHC data \cite{Rathexp_fit} and EPOSlhc event generator results, using Eq. (\ref{eq:fitttt}) in $Pb-Pb$ and $p-p$ collision systems at energies $5.02$ and $13$ TeV, respectively.}
\label{fig:ratios}
\end{figure}

 It is noticed that at 5.02 TeV for $pb-pb$ collision the theoretical results are rather matched with the experimental data compared to the fitted one in a previous study \cite{Rathexp_fit}. In the future work, we would focus on p-p collision \cite{Khutia2019} at LHC 7 TeV and high multiplicity and for $Pb-Pb$ collisions at 2.76 TeV energies \cite{Adam2015,Aamodt2011,Aad2012,Chatrchyan2011} due to its importance in the studying of the hadronic matter under extreme conditions. 
 
\begin{table}[htbp]
 \begin{center}
\caption{A set of compound particles ratios used for fitting HRG model calculations with both LHC data \cite{Rathexp_fit} and EPOSlhc event generator in $Pb-Pb$ and $p-p$ collision systems at energies $5.02$ and $13$ TeV, respectively.}
    \label{Table2}
 \begin{tabular}{c | c| c|c |c|c} 
 \hline \hline
$\frac{k^{+}+k^{-}}{\pi^{+}+\pi^{-}}$& $\frac{p+\overline{p}}{\pi^{+}+\pi^{-}}$&  $\frac{2k_{s}^{0*}}{\pi^{+}+\pi^{-}}$&  $\frac{2\Lambda}{\pi^{+}+\pi^{-}}$& $\frac{2\phi}{\pi^{+}+\pi^{-}}$&  $\frac{\Xi^{+}+\Xi^{-}}{\pi^{+}+\pi^{-}}$
\\
 \hline \hline
  \end{tabular}
\end{center}
\end{table}

 \end{itemize}

 \begin{table}[htbp]
 \begin{center}
\caption{The estimated $T_{ch}$ and $\mu_{B}$ as a result of fitting the calculated results from the HRG model of the particle ratios presented in Tab. (\ref{Table2}) with both LHC and EPOSlhc results incomparison with that shown in \cite{Rathexp_fit}.}
    \label{Table3}
    \begin{adjustbox}{width=\columnwidth}
\begin{tabular}{|c|c|c|c|c|c|c|c|c|c|} 
\hline
  \multirow{2}{*}{ $\sqrt{s} [TeV]$} & \multicolumn{3}{|c|}{LHC} & \multicolumn{3}{|c|}{EPOSlhc}&\multicolumn{3}{|c|}{\cite{Rathexp_fit}}  \\ 
\cline { 2 - 10 }
   & $T_{ch} [GeV]$ & $\mu_{B} [MeV] $ & $\chi^{2}/dof$ & $T_{ch} [GeV]$& $\mu_{B} [MeV] $ & $\chi^{2}/dof$& $T_{ch} [GeV]$& $\mu_{B} [MeV] $ & $\chi^{2}/dof$ \\ 
   \hline 
    5.02  & 0.145 $\pm$ 0.0052 & 0.3 $\pm$ 0.0026 & 68.9771/6 &0.140 $\pm$ 0.0043 & 0.35 $\pm$ 0.0031 &51.3136/6 & 0.1491$\pm$0.0021& - & 48.529/3\\
   \hline
    13 & 0.149 $\pm$0.0042& 0.25 $\pm$ 0.0025  & 78.579/6& 0.147 $\pm$ 0.0032 & 0.27 $\pm$ 0.0028 &72.903/6 &0.1579 $\pm$ 0.0023& - & 31.766/3 \\
\hline
\end{tabular}
\end{adjustbox}
\end{center}
\end{table}

The extracted chemical freeze-out temperature and baryon chemical potential are shown in Tab. (\ref{Table3}). It seems that, hadronization occurs at $135$ MeV and small $\mu_{B}$ 0.6 MeV as compared to a previous study \cite{Rathexp_fit} in which hadronization occurred at $T_{ch}$ 140-160 MeV and $\chi^{2}/dof$= 48.529/3  (\textbf{dof} :- degree of freedom ). However, in ref.\cite{Rathexp_fit} the fitted particle ratios are 3 only, the present work is closer to the minimum fitted temperature $T_{ch}\sim 140 MeV$. On the other hand, the expected critical temperature in (\cite{Bazavov2018}) is (156.5$\pm$ 1.5) MeV. 

\begin{itemize}
\item
Analysis of Particles Ratios \\
The statistical fit between the obtained results from HRG model and both LHC data and EPOSlhc event generator are done as follows 
\begin{equation}
\chi^{2} = \sum_{i} \frac{R_{i}^{exp}-R_{i}^{model}}{\sigma_{i}^{2}}.
\label{eq:fitttt}
\end{equation}
where $R_{i}^{exp}$ and $R_{i}^{model}$ represent the experimental and computed values of the particles ratios, respectively. $\sigma_{i}$ is the error in the experimental results.
\end{itemize}

  In the present work, all hadrons and resonances are implemented from the particle data group (PDG) up to 11 GeV \cite{PDG2014}. The main point in Tab. (\ref{Table3}) is that the data agree fairly well with Pb-Pb collision but shows a great discrepancy for p-p collision. This note is marked in an earlier work \cite{Sharma2019}, between the different system sizes and using the canonical and grand canonical ensemble, which was used the latter in the present work.

\section{Conclusion}\label{conclusion}
 In the present work, various well-identified, strange and multi-starnge particles ratios as a function of different centre of mass energies are studied using various models namely; HRG model, EPOS $1.99$ and EPOSlhc event generators. The aim of using both event generators to predict the particles ratios that are not measured yet. The obtained results from both event-generators and HRG model are compared with the available experimental data and show a good agreement along the whole range of the considered $\sqrt{s}$. The strangeness enhancement is studied in terms of the strange particle multiplicities and the dependence of the centre of mass energies of the mentioned particle ratios. The production of particles that contain one strange quark, or multi-strange quarks are enhanced. The ratio of strange particles is doubled from 0.5 to 1 at energies $\sqrt{s}= $ 0.001 to 13 TeV, this is clearly shown in Figs. (\ref{fig:Second}, \ref{fig:Third}). Such particles have no quarks in the colliding nuclei (Pb-Pb) or colliding nucleons (p-p). Therefore, the enhancement of these particles are considered a strong probe for the QGP formation. Particularly, the strange quarks may be produced from the deconfinement of matter phase. For more investigation of strangeness enhancement, EPOSlhc event-generator is used alongside with the HRG model for strange, non-strange and multi-strange particles, it is elaborated for Pb-Pb collision.
 
Fitting the HRG results with the experimental data has an important impact of using canonical and grand canonical ensemble, in which the latter has a global conservation of the different quantum numbers. The comparison of different ensembles has been carried out \cite{Sharma2019}, and it was concluded that there are discrepancies between the different ensembles and with the various collision system sizes such as $p-p$, $pb-pb$, and $p-pb$. This final conclusion motivates the authors to extend this work to study the different ensembles at LHC energies in TeV range.

\section{Acknowledgment}
This paper has been assigned the permanent arXiv identifier 2105.14303. The authors would like to thank the anonymous referee for the valuable comments and suggestions for future work, and improve the manuscript.

\providecommand{\href}[2]{#2}\begingroup\raggedright\endgroup

\end{document}